\documentclass[%
 reprint,%
 amssymb, amsmath,%
 aip,cha,%
]{revtex4-1}

\usepackage[colorlinks=true,linkcolor=blue,citecolor=blue,urlcolor=blue]{hyperref}%
\usepackage{docs}%
\usepackage{bm}%
\usepackage{graphicx}
\usepackage{units}
\expandafter\ifx\csname package@font\endcsname\relax\else
 \expandafter\expandafter
 \expandafter\usepackage
 \expandafter\expandafter
 \expandafter{\csname package@font\endcsname}%
\fi
\usepackage[outdir=./]{epstopdf}
\newcommand{\bea}{\begin{eqnarray}}
\newcommand{\eea}{\end{eqnarray}}
\newcommand{\be}{\begin{equation}}
\newcommand{\ee}{\end{equation}}

\newcommand{\rmg}{\mathrm{g}}

\newcommand{\rmb}{\mathrm{b}}
\newcommand{\rmf}{\mathrm{f}}

\newcommand{\rmpp}{\mathrm{p}}
\newcommand{\rmc}{\mathrm{c}}
\newcommand{\rmh}{\mathrm{h}}
\newcommand{\kBT}{k_\mathrm{B}T}

\newcommand{\rmI}{\mathrm{I}}

\begin{document}

\title{
Reversible thermal diode and energy harvester with a superconducting quantum interference single-electron transistor}%

\author{Donald Goury}%
\affiliation{Magist\`ere de Physique Fondamentale, Universit\'e Paris-Saclay, F-91405 Orsay, France\looseness=-1}%
\author{Rafael S\'anchez}%
\affiliation{Departamento de F\'isica Te\'orica de la Materia Condensada, and Condensed Matter Physics Center (IFIMAC), Universidad Aut\'onoma de Madrid, 28049 Madrid, Spain}%

\date[]{\href{https://doi.org/10.1063/1.5109100}{Appl. Phys. Lett. {\bf 115}, 092601 (2019)}}%

\begin{abstract}
The density of states of proximitized normal nanowires interrupting superconducting rings can be tuned by the magnetic flux piercing the loop. Using these as the contacts of a single-electron transistor allows to control the energetic mirror asymmetry of the conductor, this way introducing rectification properties. In particular, we show that the system works as a diode that rectifies both charge and heat currents and whose polarity can be reversed by the magnetic field and a gate voltage. We emphasize the role of dissipation at the island. The coupling to substrate phonons enhances the effect and furthermore introduces a channel for phase tunable conversion of heat exchanged with the environment into electrical current.
\end{abstract}

\maketitle


Nanometric electronic devices demand on-chip components that are driven by temperature gradients or that manage thermal flows. Indeed, nanoscale conductors are interesting in this sense because of their intrinsic nonlinearities, strong spectral features and tunability~\cite{benenti_fundamental_2017,giazotto_opportunities_2006}. However, progress in this direction has been slow due to difficulties in the precise experimental detection of heat currents.  Only very recently, great advances have been achieved in various mesoscopic configurations~\cite{giazotto_josephson_2012,martinez-perez_coherent_2014,jezouin_quantum_2013,riha_mode-selected_2015,cui_quantized_2017,dutta_thermal_2017}. They open the way to 
realize theoretical proposals of heat to current converters~\cite{hotspots,holger,cavities,hartmann,roche_harvesting_2015,
humphrey_reversible_2002,josefsson_quantum-dot_2018,resonant,gulzat}, refrigerators~\cite{courtois_electronic_2014,
edwards_quantum_1993,prance_electronic_2009,nahum_electronic_1994,leivo_efficient_1996,
saira_heat_2007,pekola_refrigerator_2014,feshchenko_experimental_2014,koski_onchip_2015}, thermal transistors~\cite{thierschmann_thermal_2015,transistor} and diodes~\cite{scheibner_quantum_2008,giazotto_thermal_2013,martinez-perez_efficient_2013,martinez-perez_rectification_2015}, and valves~\cite{ronzani_tunable_2018} in the lab.

A thermal rectifier is a system that responds to reversed temperature gradients with currents of different magnitude~\cite{benenti_from_2016}. 
For it to work as a thermal diode, forward and backward flows must be of different orders of magnitude. In electronic systems, this is the case for two terminal mesoscopic junctions with strong non-linearities due to Coulomb interactions~\cite{scheibner_quantum_2008,ruokola_thermal_2009,
sierra_nonlinear_2015,vannucci_interference_2015,tfg} or coupled to an additional thermal bath with which it exchanges energy but no charge~\cite{fornieri_normal_2014,diode,jiang_phonon_2015,guillem,devices}. The performance of the device is then controlled by an external parameter, typically a gate voltage. A requirement is the absence of mirror symmetry, which can also be introduced in the spectral properties of the contacts~\cite{giazotto_thermal_2013,martinez-perez_efficient_2013,martinez-perez_rectification_2015,oettinger_heat_2014,bours_phase-tunable2019}. 

\begin{figure}[b]
\includegraphics[width=\linewidth]{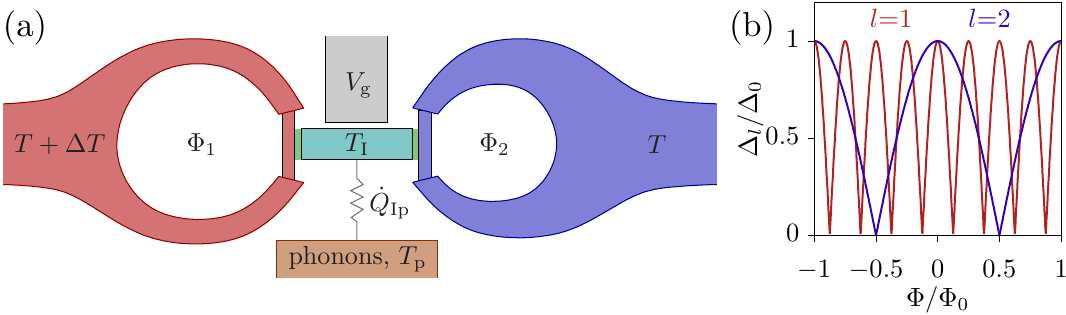}\\
\includegraphics[width=\linewidth]{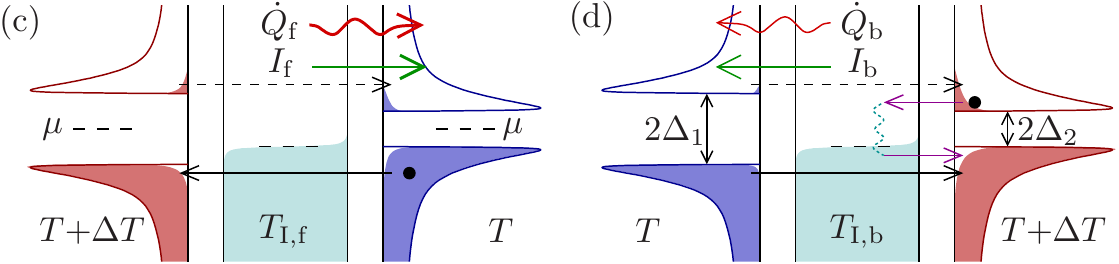}
\caption{\label{fig:scheme} SQUISET rectifier. (a) A metallic island (controlled by a gate voltage $V_\rmg$) is tunnel coupled to the weak links of two superconducting loops held at different temperatures, $T$ and $T{+}\Delta T$, but the same chemical potential $\mu$. (b) The magnetic flux piercing each loop controls the gap of the proximitized contacts, $\Delta_l$, yielding asymmetric spectra when their area is different, such that $\Phi{=}\Phi_2{\neq}\Phi_1$ (here, $\Phi_1{=}4\Phi)$. Heat exchanged with a phonon bath at temperature $T_{\rm p}$ defines the island temperature, $T_{\rm I}$. Energy filtering due to the superconducting gaps affect differently the (c) foward and (d) backward currents if $\Delta_1\neq\Delta_2$.}
\end{figure}

At low temperatures, hybrid metallic-superconducting junctions are interesting candidates
as one can make use of the properties of single-electron tunneling in strongly interacting islands~\cite{averin_coulomb_1986} and of the energy filtering introduced by the gap of the superconducting contacts. Here we investigate a superconducting quantum interference single-electron transistor (SQUISET), sketched in Fig.~\ref{fig:scheme}(a). Similar setups have recently been proposed~\cite{enrico_superconducting_2016} and implemented~\cite{enrico_phase_2017} as single-particle sources and heat valves~\cite{strambini_proximity_2014}. It consists on a normal metal island in the strong Coulomb blockade regime such that its occupation fluctuates between $n$=0,1 extra electrons. It can be controlled by the voltage $V_{\rmg}$ of a plunger gate coupled to it via a capacitance $C_\rmg$. The island is tunnel-coupled to two short wires that close two respective superconducting rings serving as contacts. Due to the proximity effect, the wires acquire a minigap~\cite{heikkila_supercurrent_2002,leSueur_phase_2008} that is controlled by the magnetic flux, $\Phi_l$, piercing the corresponding ring~\cite{enrico_superconducting_2016}:
\be
\Delta_l=\Delta_0\left|\cos{\frac{\pi\Phi_l}{\Phi_0}}\right|,
\ee
with $l$=1,2 and where $\Delta_0$ is the gap of the superconducting rings and $\Phi_0$ is the flux quantum. This way, the island is effectively coupled to contacts whose spectral properties can be tuned with an external magnetic field. Importantly for our purposes, their asymmetry can also be controlled if the size of the rings is different, as in Fig.~\ref{fig:scheme}(b). In particular, one can  tune it quite flexibly from configurations with $\Delta_1{\gg}\Delta_2$ to the opposite. This introduces the possibility to find a thermal diode whose polarity can be furthermore reverted on chip. The interplay between magnetic and electrostatic control (via $\Phi$ and $V_\rmg$, respectively) allows for an enhanced tunability of the device both for heat and thermoelectric currents responding to longitudinal temperature gradients $\Delta T$ between terminals 1 and 2. We assume they are at the same chemical potential, $\mu{=}0$.

Thermalization in the island plays a central role as, on one hand, it establishes a well defined electronic distribution and, on the other hand, it dissipates part of the heat injected from the leads. The forward and backward currents are given by the response of the barrier separating the island, at temperature $T_\rmI$, and the  corresponding cold reservoir, at temperature $T$. If $\Delta_1{\neq}\Delta_2$, the electrons with an energy $\Delta_l{<}E{<}\Delta_{l'}$ in the lead $l$ with smaller gap do not contribute to charge transport, but heat the island up, see Fig.~\ref{fig:scheme}(d). This has the double effect of (i) introducing an asymmetry between currents that leads to rectification, and of (ii) inducing a different $T_\rmI$ for the forward and backward configurations, as illustrated in Figs.~\ref{fig:scheme}(c) and (d). As discussed below, (ii) also has an impact on the rectification.

As heat dissipated at the island is important, the coupling to substrate phonons, at a temperature $T_\rmpp$, cannot be disregarded~\cite{wellstood_hot_1994}. It in turn helps to increase the rectification.
Interestingly, it also makes the system work as an energy harvester that converts heat exchanged with the phonon bath into an electric current~\cite{bjorn_review}.


The metallic island is described by the simple model~\cite{averin_coulomb_1986}
$H=E_\rmI(n-n_\rmg)^2$, with a charging energy $E_\rmI=e^2/(2C)$ given by its capacitance, $C$, and the elementary charge $e$. The charge of the island can be externally tuned via $n_{\rmg}{=}C_\rmg V_\rmg/e$.
Sequential tunneling events through terminal $l$ into the island are described by the rates:
\begin{align}
\label{rates}
\Gamma_{l}^{+}{=}\frac{1}{e^2R_l}\int {dE}{\cal N}_{l}(E)f_l(E)[1{-}f_\rmI(E{-}U)]
\end{align}
with the associated resistances $R_l$~\cite{freq}, and the Fermi functions $f_i(E)=\left(1{+}e^{E/\kBT_i}\right)^{-1}$. The difference of chemical potentials for this transition is $U=E_\rmI\left(1-2n_\rmg\right)$.
This defines the charge degeneracy point at $n_\rmg=1/2$.
Tunneling out rates, $\Gamma_{l}^{-}$, are obtained by replacing $f_i(E)\to1-f_i(E)$ in Eq.~\eqref{rates}. 
At finite temperatures, subgap states appear in the superconducting leads, which are taken into account by the Dynes density of states:~\cite{pekola_environment_2010}
\be 
{\cal N}_{l}(E)=\left|{\rm Re}\left(\frac{E+i\gamma}{\sqrt{(E+i\gamma)^2-\Delta_l^2}}\right)\right|,
\ee
where $\gamma$ is the quasiparticle lifetime broadening.


With the particle tunneling rates $\Gamma_{l}^\pm$ one obtains the steady state occupation of the island $p_n$ via the stationary solution of the rate equations: 
\be
\dot p_0=-\dot p_1=\sum_l\left(\Gamma_l^-p_1-\Gamma_l^+p_0\right),
\ee
with $p_0+p_1=1$.
The charge and heat currents into $l$ are hence respectively given by:
\begin{align}
\label{curr}
I_l&=-e\left[\Gamma_{l}^-p_1-\Gamma_{l}^+p_0\right]\\
\dot{Q}_l&=\tilde\Gamma_{l}^-p_1-\tilde\Gamma_{l}^+p_0,
\end{align}
where  $\tilde\Gamma_l^\pm$ are obtained by inserting $E$ in the integral of the corresponding $\Gamma_l^\pm$. As no Joule heating is generated in the system, energy conservation is maintained by the heat absorbed at the island: $\dot{Q}_\rmI=-\dot{Q}_1-\dot{Q}_2$. The temperature of the island is then given by the balance of the heat injected from the leads and the heat exchange with the phononic environment via the equation: 
\be
\label{eq:balance}
\dot Q_{\rmI}(T_\rmI)=\lambda(T_{{\rm p}}^5-T_\rmI^5),
\ee
where $\lambda$ depends on the dimensions of the island and its material~\cite{giazotto_opportunities_2006}. At the superconducting contacts, this exchange is suppressed by the gap~\cite{timofeev_recombination_2009}.

\begin{figure}[t]
\includegraphics[width=\linewidth]{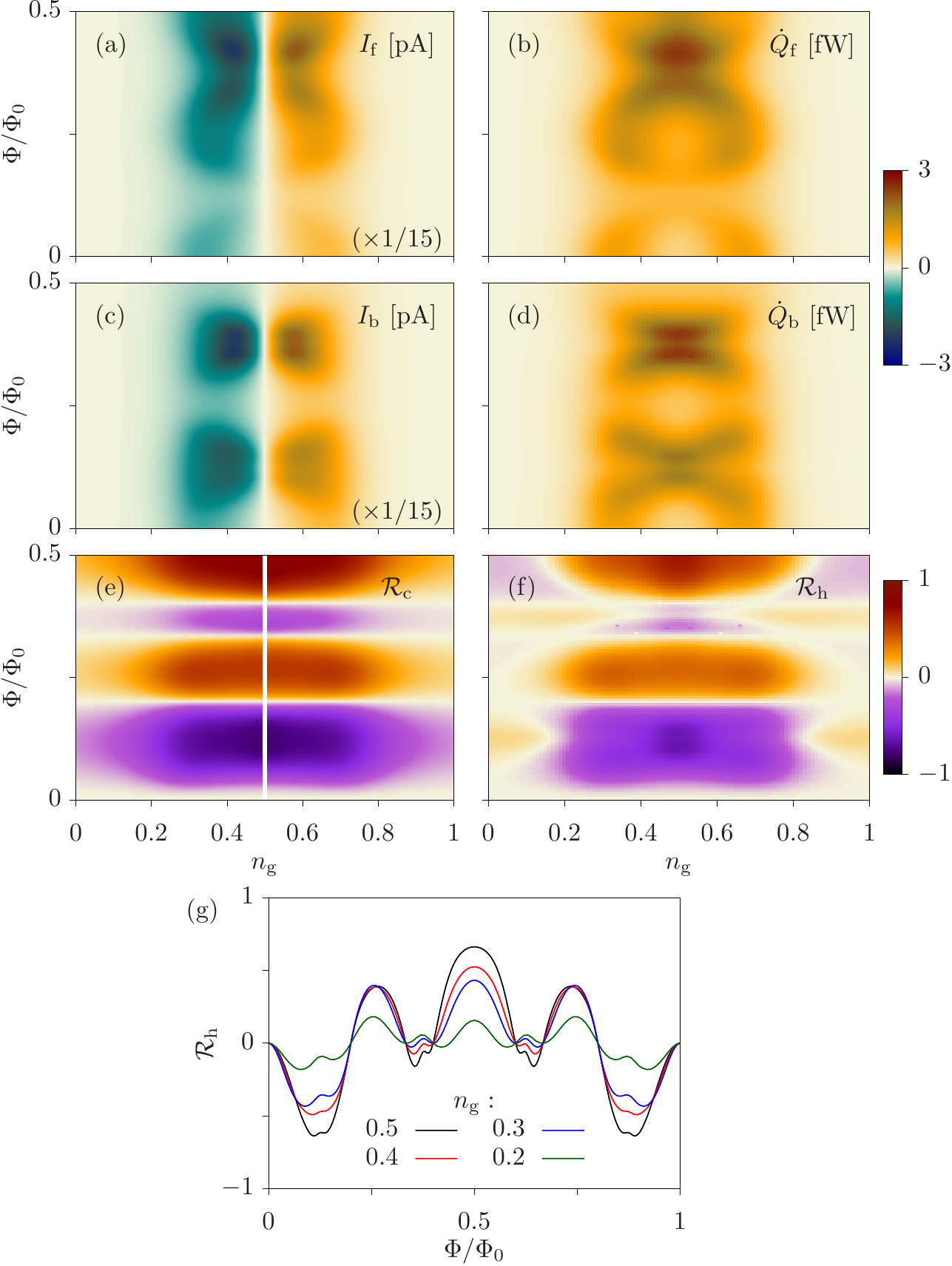}
\caption{\label{fig:magnet} Forward (a,b) and backward (c,d) charge and heat currents as functions of the gating, $n_{\rm g}$, and the magnetic field, $\Phi{=}\Phi_2{=}\Phi_1/4$. The typical sign change of thermoelectric currents and double peak structure of heat currents around $n_\rmg{=}1/2$ of single-electron transistors is modulated by the magnetic field. (e,f) Rectification coefficients for the corresponding currents. They change sign when $\Delta_1{=}\Delta_2$. (g) Cuts of ${\cal R}_\rmh$ in (f) for different $n_\rmg$ show its change of polarity with gating and flux. Parameters: $E_\rmI{=}\unit[0.25]{meV}$, $R_1{=}R_2{=}R_0{=}\unit[100]{k\Omega}$, $T{=}\unit[116]{mK}$, $\Delta T{=}2T$, $T_\rmpp{=}\unit[58]{mK}$, $\lambda{=}\lambda_0{=}\unit[0.45]{meV^2K^{-5}}$, and $\gamma{=}\unit[10^{-3}]{meV}$.}
\end{figure}

The diode performance is characterized by a difference of the currents under opposite thermal bias. We thus write the forward, $I_\rmf=I_2(T_1{=}T{+}\Delta T,T_2{=}T)$, and backward, $I_\rmb=I_1(T_1{=}T,T_2{=}T{+}\Delta T)$, charge currents, and accordingly for the heat currents $\dot{Q}_\rmf$ and $\dot{Q}_\rmb$. With these, we define the rectification coefficients:
\be
\label{eq:rectI}
{\cal R}_\rmc=\frac{|I_\rmf|-|I_\rmb|}{{\rm max}(|I_\rmf|,|I_\rmb|)} \text{  and  } {\cal R}_\rmh=\frac{|\dot{Q}_\rmf|-|\dot{Q}_\rmb|}{{\rm max}(|\dot{Q}_\rmf|,|\dot{Q}_\rmb|)}.
\ee
Note that the sign of ${\cal R}_\alpha$ informs of the polarity of the diode, but this is not necessarilly related to the sign of the currents. In particular, for the charge currents, both forward and backward flows change sign around the electron-hole symmetric configurations with $n_\rmg=1/2$, as expected for thermoelectric currents in Coulomb blockade systems~\cite{staring_coulomb-blockade_1993,dzurak_observation_1993,beenakker_theory_1992,erdman_thermoelectric_2017}, see Fig.~\ref{fig:magnet}. However their amplitude depends on the gap of the contacts, and hence ${\cal R}_\rmc$ oscillates and changes sign with the magnetic flux, vanishing when $\Delta_1=\Delta_2$, see Figs.~\ref{fig:magnet}(a), (c) and (e). On the other hand, if the phonons do not introduce an additional thermal gradient, $\dot{Q}_\rmf$ and $\dot{Q}_\rmb$ are always positive, see Figs.~\ref{fig:magnet}(b) and (d). Note that due to heat flowing into the island, ${\cal R}_\rmh$  also changes its polarity with $n_\rmg$, as shown in Figs.~\ref{fig:magnet}(f) and (g). This is not the case for ${\cal R}_\rmc$ due to charge conservation in the leads ($I_1+I_2=0$).

The maximal rectification coefficients are found close to the electron-hole symmetric configurations. However, additional maxima appear by tuning $n_g$ as an effect of the border of the superconducting gap. Note that ${\cal R}_\rmc$ is not well defined at $n_\rmg{=}1/2$, as no thermoelectric currents can be generated in electron-hole symmetric configurations.

\begin{figure}[t]
\includegraphics[width=\linewidth]{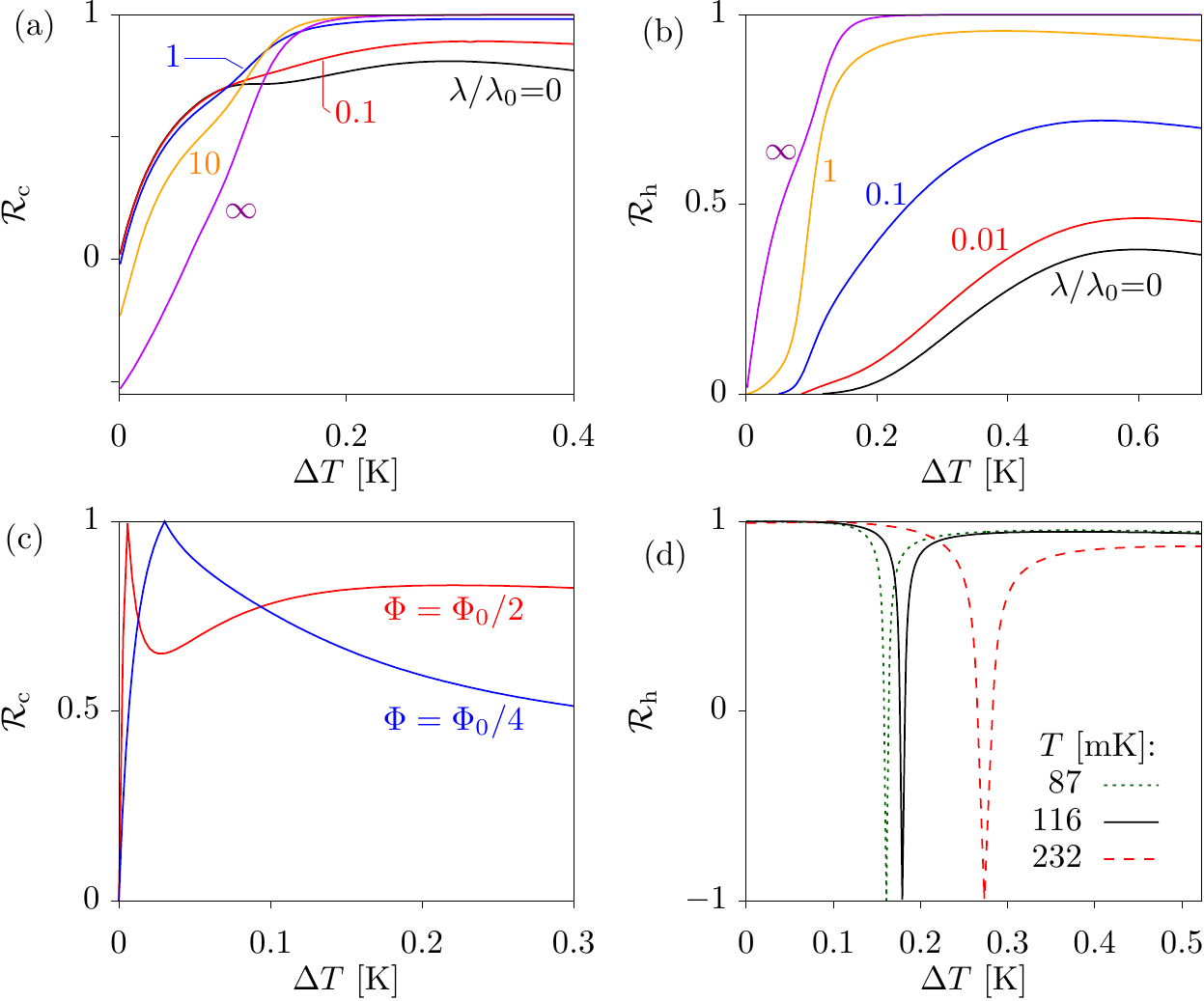}
\caption{\label{fig:rectif} Effect of the phonon bath. (a,b) Effect of the coupling to the phonon bath, $\lambda$, on the charge and heat rectification coefficients, with $T{=}T_\rmpp{=}\unit[58]{mK}$. $T_\rmI$ is given by the balance Eq.~\eqref{eq:balance}. (c) Charge and (d) heat rectification coefficients for fixed phonon temperature $T_\rmpp{=}\unit[58]{mK}$. In (c) $n_\rmg$ is chosen such that $I_\rmf$ is maximized for different magnetic fields: $\Phi{=}\Phi_0/2$ (with $n_\rmg{=}0.49$), and $\Phi{=}\Phi_0/4$ (with $n_\rmg{=}0.3349$). In all cases, except when mentioned, parameters are as in Fig.~\ref{fig:magnet}, with magnetic fields are fixed such that $\Delta_1{=}\unit[0.193]{meV}$ and $\Delta_2{=}\unit[0.01]{meV}$, $n_\rmg{=}0.49$ and $\gamma{=}\unit[10^{-5}]{meV}$.}
\end{figure}

We now investigate the crucial effect of the phonon bath. Let us consider first the case with $T_\rmpp=T$, see Figs.~\ref{fig:rectif}(a) and (b). If the coupling to the phonons is strong, the island dissipates the incoming heat from the leads into the environment, so its temperature $T_\rmI$ is not affected (as if it was connected to a thermostat). Then for low $\Delta T$, the gap suppresses the forward current (in this configuration with $\Delta_1>\Delta_2$), giving ${\cal R}_\rmc<0$. Increasing the gradient, states over the gap are populated such that $I_\rmf$ and $I_\rmb$ cross, and ${\cal R}_\rmc$ changes sign.
The rectification effects are then enhanced and high coefficients ${\cal R}_\rmc$ and ${\cal R}_\rmh$ are achieved, see Figs.~\ref{fig:rectif}(a) and (b). 
With smaller $\lambda$, the temperature of the island increases, which is detrimental for the diode performance and avoids the change of sign of ${\cal R}_\rmc$. 
The optimal values of $\cal R_\alpha$ are no affected by quasiparticle lifetimes in the experimentally relevant range considered here (with $\gamma\sim10^{-4}$--$10^{-5}$~meV). 

In real configurations, electrons and phonons are not necessarily at the same temperature. The presence of an additional temperature gradient $T_\rmpp{-}T$ introduces unexpected effects in the rectification properties of the system, due the local breaking of detailed balance. This introduces crossed thermoelectric effects: the phonon bath behaves as a third terminal with which the electronic system exchanges heat (and of course, no charge). This heat exchange can be converted into a contribution of the charge current whose sign is defined by both mirror and electron-hole asymmetries~\cite{entinwWohlman_three-terminal_2010,banff}, even if $\Delta T=0$. This additional current can flow in the opposite direction as the backward current and eventually cancel it. At that point, only the forward current is finite and we get ${\cal R}_\rmc=1$, as shown in Fig.~\ref{fig:rectif}(c) for two different configurations of the magnetic flux and the gate voltage.

The behaviour of the heat currents is different, see Fig.~\ref{fig:rectif}(d). At low temperature gradients $\Delta T$, heat injected from the phonon bath dominates and reverts the forward current, while the backward one is strongly suppressed (due to the gap $\Delta_1$). This leads to ${\cal R}_\rmh{\approx}1$. As $\Delta T$ increases, the longitudinal heat current dominates, so the forward flow changes sign. At that point, as only the forward current flows, ${\cal R}_{\rmh}{\approx}-1$. The position of this sharp feature changes with the temperature of the electrons.

\begin{figure}[t]
\includegraphics[width=0.45\linewidth]{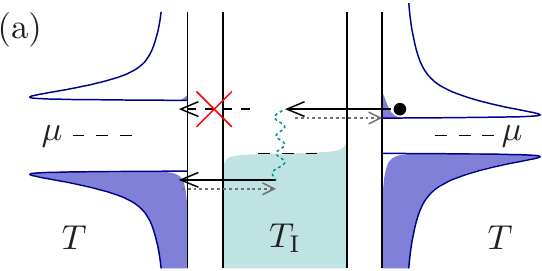}\\
\vspace{0.2cm}
\includegraphics[width=\linewidth]{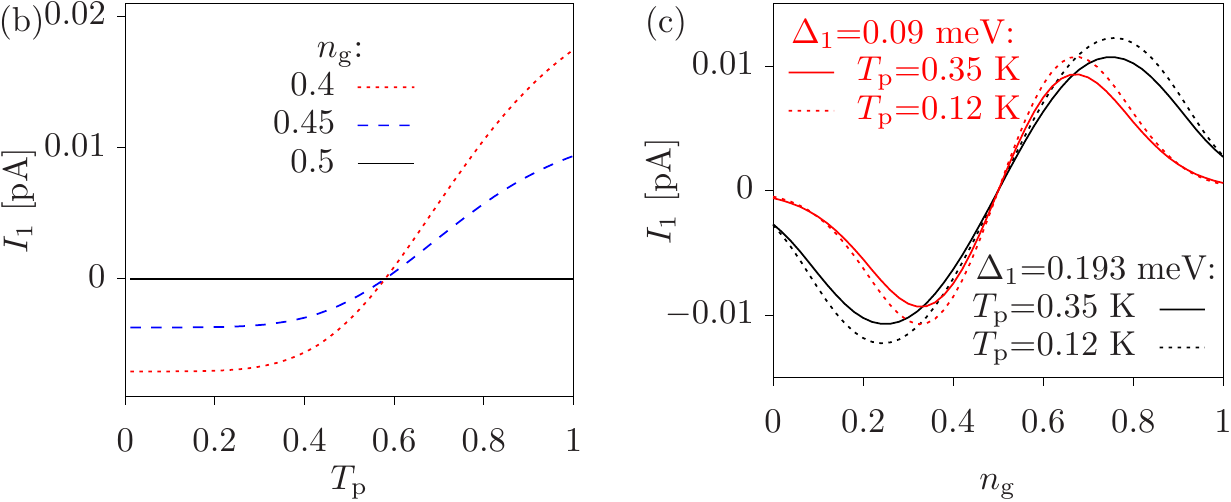}
\caption{\label{fig:harvest} Conversion of heat exchanged with the phonons. 
(a) At subgap energies, tunneling is avoided in one terminal (here, the left one due to its larger gap) but not in the right one. The sketched transition hence contributes to transport from terminal 2 to 1 mediated by thermalization at the island. The reversed sequence (marked by dotted grey arrows) is suppressed by broken local detailed balance if $T\neq T_\rmI$ and $n_\rmg\neq1/2$, leading to a finite current, $I_1$. (b) It increases with the temperature difference $T-T_\rmpp$ and can be tuned with the gate voltage (again for fixed $\Delta_1=\unit[0.193]{meV}$, $\Delta_2=\unit[0.01]{meV}$). (c) The gating parameter controls the electron-hole asymmetry and hence the sign of the current. The position of the maximum current increases with the gap $\Delta_1$ (here fixing $\Delta_2=\unit[0.01]{meV}$). In these plots, $T=\unit[58]{mK}$ and $R_0=\unit[10]{k\Omega}$.}
\end{figure}

Let us finally further explore the charge current generated by the temperature difference with the phonon bath. For this, we consider the case $T_1=T_2=T$ such that there is no longitudinal thermoelectric effect. The conductor is in equilibrium, except for being coupled to phonons at the island. The electrons in the island thermalize at a temperature given by  Eq.~\eqref{eq:balance}. Thus a local temperature difference is established in the electronic system: In the likely case that $T_\rmpp<T$, the island hence becomes a cold spot, with $T_\rmI<T$. The effect of having broken electron-hole symmetry (for $n_\rmg\neq1/2$) and broken mirror symmetry (due to the superconducting gaps) breaks local detailed balance at both barriers, i.e. $\Gamma_l^+p_0\neq\Gamma_l^-p_1$~\cite{banff}. According to Eq.~\eqref{curr}, a net charge current is generated, as sketched in Fig.~\ref{fig:harvest}(a) and shown if Fig.~\ref{fig:harvest}(b). In this sense, it is a single-electron analogue of a thermocouple in which the electron-hole asymmetry is externally tunable by means of the magnetic field. Contrary to usual thermoelectric heat engines, where heat from a hot bath is transformed into power, the generated charge current is here due to heat emitted into a colder environment. Furthermore, the sign of the current can in this case be controlled either by tuning the gate voltage or the magnetic flux, see Fig.~\ref{fig:harvest}(c). As a function of $n_\rmg$, the current has a maximum when the chemical potential of the island is close to the gap, such that the asymmetry is larger. For higher chemical potentials the effect of the superconductor is reduced and the current is suppressed. 

We emphasize that this way, the system has a dual performance as a diode and as a three-terminal converter of environmental heat~\cite{bjorn_review}. 
Differently from all-normal devices~\cite{entinwWohlman_three-terminal_2010,bosisio_nanowire_2016,jiang_thermoelectric_2012}, the presence of the superconductors, apart from introducing the required asymmetric energy filtering, ensures that heat is exchanged with the phonon bath at the mesoscopic region (the island, in our case) only: in the leads, it is strongly suppressed by the gap. This is expected to increase the efficiency. As the temperature difference with the phonons is in principle small, and so are the tunneling couplings, we do not expect the system to be a powerful heat converter. However, this effect can be used as a probe of the environmental temperature.

To conclude, we have investigated the response of a superconducting quantum interference single-electron transistor to temperature gradients. The interplay of Coulomb interactions in the island and phase coherence in the superconducting rings leads to the rectification of both charge and heat currents. The system then works as a phase and gate tunable thermal diode whose polarity can furthermore be reversed in various configurations. We find that electron-phonon coupling at the metallic island plays an essential role, with heat exchanged with the phonon bath enhancing the diode performance, which is comparable to existing experiments on related configurations~\cite{martinez-perez_rectification_2015}. Transport is sensitive to temperature differences with the phonons. This leads to salient features such as the vanishing of forward or backward flows, resulting in optimal diode behaviours with ${\cal R}_{\rmc/\rmh}=\pm1$, or in the generation of charge currents in a conductor which is otherwise in equilibrium (with $T_1=T_2$). The device is then proposed as a versatile and efficient element for heat management in the nanoscale.

We thank discussions and comments from A. Levy Yeyati, C. Urbina and F. Giazotto. This work was supported by the Spanish Ministerio de Econom\'ia, Industria y Competitividad (MINECO) via the Ram\'on y Cajal program RYC-2016-20778, and the ``Mar\'ia de Maeztu” Programme for Units of Excellence in R\&D (MDM-2014-0377). We also acknowledge the Universit\'e Paris-Saclay international grants, the EU Erasmus program.

\bibliographystyle{apsrev4-1}
\bibliography{papersquipt}

\begin{thebibliography}{64}%
\makeatletter
\providecommand \@ifxundefined [1]{%
 \@ifx{#1\undefined}
}%
\providecommand \@ifnum [1]{%
 \ifnum #1\expandafter \@firstoftwo
 \else \expandafter \@secondoftwo
 \fi
}%
\providecommand \@ifx [1]{%
 \ifx #1\expandafter \@firstoftwo
 \else \expandafter \@secondoftwo
 \fi
}%
\providecommand \natexlab [1]{#1}%
\providecommand \enquote  [1]{``#1''}%
\providecommand \bibnamefont  [1]{#1}%
\providecommand \bibfnamefont [1]{#1}%
\providecommand \citenamefont [1]{#1}%
\providecommand \href@noop [0]{\@secondoftwo}%
\providecommand \href [0]{\begingroup \@sanitize@url \@href}%
\providecommand \@href[1]{\@@startlink{#1}\@@href}%
\providecommand \@@href[1]{\endgroup#1\@@endlink}%
\providecommand \@sanitize@url [0]{\catcode `\\12\catcode `\$12\catcode
  `\&12\catcode `\#12\catcode `\^12\catcode `\_12\catcode `\%12\relax}%
\providecommand \@@startlink[1]{}%
\providecommand \@@endlink[0]{}%
\providecommand \url  [0]{\begingroup\@sanitize@url \@url }%
\providecommand \@url [1]{\endgroup\@href {#1}{\urlprefix }}%
\providecommand \urlprefix  [0]{URL }%
\providecommand \Eprint [0]{\href }%
\providecommand \doibase [0]{http://dx.doi.org/}%
\providecommand \selectlanguage [0]{\@gobble}%
\providecommand \bibinfo  [0]{\@secondoftwo}%
\providecommand \bibfield  [0]{\@secondoftwo}%
\providecommand \translation [1]{[#1]}%
\providecommand \BibitemOpen [0]{}%
\providecommand \bibitemStop [0]{}%
\providecommand \bibitemNoStop [0]{.\EOS\space}%
\providecommand \EOS [0]{\spacefactor3000\relax}%
\providecommand \BibitemShut  [1]{\csname bibitem#1\endcsname}%
\let\auto@bib@innerbib\@empty
\bibitem [{\citenamefont {Benenti}\ \emph {et~al.}(2017)\citenamefont
  {Benenti}, \citenamefont {Casati}, \citenamefont {Saito},\ and\ \citenamefont
  {Whitney}}]{benenti_fundamental_2017}%
  \BibitemOpen
  \bibfield  {author} {\bibinfo {author} {\bibfnamefont {G.}~\bibnamefont
  {Benenti}}, \bibinfo {author} {\bibfnamefont {G.}~\bibnamefont {Casati}},
  \bibinfo {author} {\bibfnamefont {K.}~\bibnamefont {Saito}}, \ and\ \bibinfo
  {author} {\bibfnamefont {R.~S.}\ \bibnamefont {Whitney}},\ }\href {\doibase
  10.1016/j.physrep.2017.05.008} {\bibfield  {journal} {\bibinfo  {journal}
  {Phys. Rep.}\ }\textbf {\bibinfo {volume} {694}},\ \bibinfo {pages} {1}
  (\bibinfo {year} {2017})}\BibitemShut {NoStop}%
\bibitem [{\citenamefont {Giazotto}\ \emph {et~al.}(2006)\citenamefont
  {Giazotto}, \citenamefont {Heikkil\"a}, \citenamefont {Luukanen},
  \citenamefont {Savin},\ and\ \citenamefont
  {Pekola}}]{giazotto_opportunities_2006}%
  \BibitemOpen
  \bibfield  {author} {\bibinfo {author} {\bibfnamefont {F.}~\bibnamefont
  {Giazotto}}, \bibinfo {author} {\bibfnamefont {T.~T.}\ \bibnamefont
  {Heikkil\"a}}, \bibinfo {author} {\bibfnamefont {A.}~\bibnamefont
  {Luukanen}}, \bibinfo {author} {\bibfnamefont {A.~M.}\ \bibnamefont {Savin}},
  \ and\ \bibinfo {author} {\bibfnamefont {J.~P.}\ \bibnamefont {Pekola}},\
  }\href {\doibase 10.1103/RevModPhys.78.217} {\bibfield  {journal} {\bibinfo
  {journal} {Rev. Mod. Phys.}\ }\textbf {\bibinfo {volume} {78}},\ \bibinfo
  {pages} {217} (\bibinfo {year} {2006})}\BibitemShut {NoStop}%
\bibitem [{\citenamefont {Giazotto}\ and\ \citenamefont
  {Mart{\ifmmode\acute{\imath}\else\'{\i}\fi}nez-P{\ifmmode\acute{e}\else\'{e}\fi}rez}(2012)}]{giazotto_josephson_2012}%
  \BibitemOpen
  \bibfield  {author} {\bibinfo {author} {\bibfnamefont {F.}~\bibnamefont
  {Giazotto}}\ and\ \bibinfo {author} {\bibfnamefont {M.~J.}\ \bibnamefont
  {Mart{\ifmmode\acute{\imath}\else\'{\i}\fi}nez-P{\ifmmode\acute{e}\else\'{e}\fi}rez}},\
  }\href {\doibase 10.1038/nature11702} {\bibfield  {journal} {\bibinfo
  {journal} {Nature}\ }\textbf {\bibinfo {volume} {492}},\ \bibinfo {pages}
  {401} (\bibinfo {year} {2012})}\BibitemShut {NoStop}%
\bibitem [{\citenamefont
  {Mart{\ifmmode\acute{\imath}\else\'{\i}\fi}nez-P{\ifmmode\acute{e}\else\'{e}\fi}rez}\
  \emph {et~al.}(2014)\citenamefont
  {Mart{\ifmmode\acute{\imath}\else\'{\i}\fi}nez-P{\ifmmode\acute{e}\else\'{e}\fi}rez},
  \citenamefont {Solinas},\ and\ \citenamefont
  {Giazotto}}]{martinez-perez_coherent_2014}%
  \BibitemOpen
  \bibfield  {author} {\bibinfo {author} {\bibfnamefont {M.~J.}\ \bibnamefont
  {Mart{\ifmmode\acute{\imath}\else\'{\i}\fi}nez-P{\ifmmode\acute{e}\else\'{e}\fi}rez}},
  \bibinfo {author} {\bibfnamefont {P.}~\bibnamefont {Solinas}}, \ and\
  \bibinfo {author} {\bibfnamefont {F.}~\bibnamefont {Giazotto}},\ }\href
  {\doibase 10.1007/s10909-014-1132-6} {\bibfield  {journal} {\bibinfo
  {journal} {J. Low Temp. Phys.}\ }\textbf {\bibinfo {volume} {175}},\ \bibinfo
  {pages} {813} (\bibinfo {year} {2014})}\BibitemShut {NoStop}%
\bibitem [{\citenamefont {Jezouin}\ \emph {et~al.}(2013)\citenamefont
  {Jezouin}, \citenamefont {Parmentier}, \citenamefont {Anthore}, \citenamefont
  {Gennser}, \citenamefont {Cavanna}, \citenamefont {Jin},\ and\ \citenamefont
  {Pierre}}]{jezouin_quantum_2013}%
  \BibitemOpen
  \bibfield  {author} {\bibinfo {author} {\bibfnamefont {S.}~\bibnamefont
  {Jezouin}}, \bibinfo {author} {\bibfnamefont {F.~D.}\ \bibnamefont
  {Parmentier}}, \bibinfo {author} {\bibfnamefont {A.}~\bibnamefont {Anthore}},
  \bibinfo {author} {\bibfnamefont {U.}~\bibnamefont {Gennser}}, \bibinfo
  {author} {\bibfnamefont {A.}~\bibnamefont {Cavanna}}, \bibinfo {author}
  {\bibfnamefont {Y.}~\bibnamefont {Jin}}, \ and\ \bibinfo {author}
  {\bibfnamefont {F.}~\bibnamefont {Pierre}},\ }\href {\doibase
  10.1126/science.1241912} {\bibfield  {journal} {\bibinfo  {journal}
  {Science}\ }\textbf {\bibinfo {volume} {342}},\ \bibinfo {pages} {601}
  (\bibinfo {year} {2013})}\BibitemShut {NoStop}%
\bibitem [{\citenamefont {Riha}\ \emph {et~al.}(2015)\citenamefont {Riha},
  \citenamefont {Miechowski}, \citenamefont {Buchholz}, \citenamefont
  {Chiatti}, \citenamefont {Wieck}, \citenamefont {Reuter},\ and\ \citenamefont
  {Fischer}}]{riha_mode-selected_2015}%
  \BibitemOpen
  \bibfield  {author} {\bibinfo {author} {\bibfnamefont {C.}~\bibnamefont
  {Riha}}, \bibinfo {author} {\bibfnamefont {P.}~\bibnamefont {Miechowski}},
  \bibinfo {author} {\bibfnamefont {S.~S.}\ \bibnamefont {Buchholz}}, \bibinfo
  {author} {\bibfnamefont {O.}~\bibnamefont {Chiatti}}, \bibinfo {author}
  {\bibfnamefont {A.~D.}\ \bibnamefont {Wieck}}, \bibinfo {author}
  {\bibfnamefont {D.}~\bibnamefont {Reuter}}, \ and\ \bibinfo {author}
  {\bibfnamefont {S.~F.}\ \bibnamefont {Fischer}},\ }\href {\doibase
  10.1063/1.4908052} {\bibfield  {journal} {\bibinfo  {journal} {Appl. Phys.
  Lett.}\ }\textbf {\bibinfo {volume} {106}},\ \bibinfo {pages} {083102}
  (\bibinfo {year} {2015})}\BibitemShut {NoStop}%
\bibitem [{\citenamefont {Cui}\ \emph {et~al.}(2017)\citenamefont {Cui},
  \citenamefont {Jeong}, \citenamefont {Hur}, \citenamefont {Matt},
  \citenamefont {Kl{\ifmmode\ddot{o}\else\"{o}\fi}ckner}, \citenamefont
  {Pauly}, \citenamefont {Nielaba}, \citenamefont {Cuevas}, \citenamefont
  {Meyhofer},\ and\ \citenamefont {Reddy}}]{cui_quantized_2017}%
  \BibitemOpen
  \bibfield  {author} {\bibinfo {author} {\bibfnamefont {L.}~\bibnamefont
  {Cui}}, \bibinfo {author} {\bibfnamefont {W.}~\bibnamefont {Jeong}}, \bibinfo
  {author} {\bibfnamefont {S.}~\bibnamefont {Hur}}, \bibinfo {author}
  {\bibfnamefont {M.}~\bibnamefont {Matt}}, \bibinfo {author} {\bibfnamefont
  {J.~C.}\ \bibnamefont {Kl{\ifmmode\ddot{o}\else\"{o}\fi}ckner}}, \bibinfo
  {author} {\bibfnamefont {F.}~\bibnamefont {Pauly}}, \bibinfo {author}
  {\bibfnamefont {P.}~\bibnamefont {Nielaba}}, \bibinfo {author} {\bibfnamefont
  {J.~C.}\ \bibnamefont {Cuevas}}, \bibinfo {author} {\bibfnamefont
  {E.}~\bibnamefont {Meyhofer}}, \ and\ \bibinfo {author} {\bibfnamefont
  {P.}~\bibnamefont {Reddy}},\ }\href {\doibase 10.1126/science.aam6622}
  {\bibfield  {journal} {\bibinfo  {journal} {Science}\ }\textbf {\bibinfo
  {volume} {355}},\ \bibinfo {pages} {1192} (\bibinfo {year}
  {2017})}\BibitemShut {NoStop}%
\bibitem [{\citenamefont {Dutta}\ \emph {et~al.}(2017)\citenamefont {Dutta},
  \citenamefont {Peltonen}, \citenamefont {Antonenko}, \citenamefont {Meschke},
  \citenamefont {Skvortsov}, \citenamefont {Kubala}, \citenamefont
  {K{\ifmmode\ddot{o}\else\"{o}\fi}nig}, \citenamefont {Winkelmann},
  \citenamefont {Courtois},\ and\ \citenamefont {Pekola}}]{dutta_thermal_2017}%
  \BibitemOpen
  \bibfield  {author} {\bibinfo {author} {\bibfnamefont {B.}~\bibnamefont
  {Dutta}}, \bibinfo {author} {\bibfnamefont {J.~T.}\ \bibnamefont {Peltonen}},
  \bibinfo {author} {\bibfnamefont {D.~S.}\ \bibnamefont {Antonenko}}, \bibinfo
  {author} {\bibfnamefont {M.}~\bibnamefont {Meschke}}, \bibinfo {author}
  {\bibfnamefont {M.~A.}\ \bibnamefont {Skvortsov}}, \bibinfo {author}
  {\bibfnamefont {B.}~\bibnamefont {Kubala}}, \bibinfo {author} {\bibfnamefont
  {J.}~\bibnamefont {K{\ifmmode\ddot{o}\else\"{o}\fi}nig}}, \bibinfo {author}
  {\bibfnamefont {C.~B.}\ \bibnamefont {Winkelmann}}, \bibinfo {author}
  {\bibfnamefont {H.}~\bibnamefont {Courtois}}, \ and\ \bibinfo {author}
  {\bibfnamefont {J.~P.}\ \bibnamefont {Pekola}},\ }\href {\doibase
  10.1103/PhysRevLett.119.077701} {\bibfield  {journal} {\bibinfo  {journal}
  {Phys. Rev. Lett.}\ }\textbf {\bibinfo {volume} {119}},\ \bibinfo {pages}
  {077701} (\bibinfo {year} {2017})}\BibitemShut {NoStop}%
\bibitem [{\citenamefont {S\'anchez}\ and\ \citenamefont
  {B\"uttiker}(2011)}]{hotspots}%
  \BibitemOpen
  \bibfield  {author} {\bibinfo {author} {\bibfnamefont {R.}~\bibnamefont
  {S\'anchez}}\ and\ \bibinfo {author} {\bibfnamefont {M.}~\bibnamefont
  {B\"uttiker}},\ }\href {\doibase 10.1103/PhysRevB.83.085428} {\bibfield
  {journal} {\bibinfo  {journal} {Phys. Rev. B}\ }\textbf {\bibinfo {volume}
  {83}},\ \bibinfo {pages} {085428} (\bibinfo {year} {2011})}\BibitemShut
  {NoStop}%
\bibitem [{\citenamefont {{Thierschmann}}\ \emph {et~al.}(2015)\citenamefont
  {{Thierschmann}}, \citenamefont {{S{\'a}nchez}}, \citenamefont {{Sothmann}},
  \citenamefont {{Arnold}}, \citenamefont {{Heyn}}, \citenamefont {{Hansen}},
  \citenamefont {{Buhmann}},\ and\ \citenamefont {{Molenkamp}}}]{holger}%
  \BibitemOpen
  \bibfield  {author} {\bibinfo {author} {\bibfnamefont {H.}~\bibnamefont
  {{Thierschmann}}}, \bibinfo {author} {\bibfnamefont {R.}~\bibnamefont
  {{S{\'a}nchez}}}, \bibinfo {author} {\bibfnamefont {B.}~\bibnamefont
  {{Sothmann}}}, \bibinfo {author} {\bibfnamefont {F.}~\bibnamefont
  {{Arnold}}}, \bibinfo {author} {\bibfnamefont {C.}~\bibnamefont {{Heyn}}},
  \bibinfo {author} {\bibfnamefont {W.}~\bibnamefont {{Hansen}}}, \bibinfo
  {author} {\bibfnamefont {H.}~\bibnamefont {{Buhmann}}}, \ and\ \bibinfo
  {author} {\bibfnamefont {L.~W.}\ \bibnamefont {{Molenkamp}}},\ }\href
  {\doibase 10.1038/nnano.2015.176} {\bibfield  {journal} {\bibinfo  {journal}
  {Nature Nanotechnology}\ }\textbf {\bibinfo {volume} {10}},\ \bibinfo {pages}
  {854} (\bibinfo {year} {2015})}\BibitemShut {NoStop}%
\bibitem [{\citenamefont {Sothmann}\ \emph {et~al.}(2012)\citenamefont
  {Sothmann}, \citenamefont {S\'anchez}, \citenamefont {Jordan},\ and\
  \citenamefont {B\"uttiker}}]{cavities}%
  \BibitemOpen
  \bibfield  {author} {\bibinfo {author} {\bibfnamefont {B.}~\bibnamefont
  {Sothmann}}, \bibinfo {author} {\bibfnamefont {R.}~\bibnamefont {S\'anchez}},
  \bibinfo {author} {\bibfnamefont {A.~N.}\ \bibnamefont {Jordan}}, \ and\
  \bibinfo {author} {\bibfnamefont {M.}~\bibnamefont {B\"uttiker}},\ }\href
  {\doibase 10.1103/PhysRevB.85.205301} {\bibfield  {journal} {\bibinfo
  {journal} {Phys. Rev. B}\ }\textbf {\bibinfo {volume} {85}},\ \bibinfo
  {pages} {205301} (\bibinfo {year} {2012})}\BibitemShut {NoStop}%
\bibitem [{\citenamefont {Hartmann}\ \emph {et~al.}(2015)\citenamefont
  {Hartmann}, \citenamefont {Pfeffer}, \citenamefont
  {H{\ifmmode\ddot{o}\else\"{o}\fi}fling}, \citenamefont {Kamp},\ and\
  \citenamefont {Worschech}}]{hartmann}%
  \BibitemOpen
  \bibfield  {author} {\bibinfo {author} {\bibfnamefont {F.}~\bibnamefont
  {Hartmann}}, \bibinfo {author} {\bibfnamefont {P.}~\bibnamefont {Pfeffer}},
  \bibinfo {author} {\bibfnamefont {S.}~\bibnamefont
  {H{\ifmmode\ddot{o}\else\"{o}\fi}fling}}, \bibinfo {author} {\bibfnamefont
  {M.}~\bibnamefont {Kamp}}, \ and\ \bibinfo {author} {\bibfnamefont
  {L.}~\bibnamefont {Worschech}},\ }\href {\doibase
  10.1103/PhysRevLett.114.146805} {\bibfield  {journal} {\bibinfo  {journal}
  {Phys. Rev. Lett.}\ }\textbf {\bibinfo {volume} {114}},\ \bibinfo {pages}
  {146805} (\bibinfo {year} {2015})}\BibitemShut {NoStop}%
\bibitem [{\citenamefont {{B. Roche}}\ \emph {et~al.}(2015)\citenamefont {{B.
  Roche}}, \citenamefont {{P. Roulleau}}, \citenamefont {{T. Jullien}},
  \citenamefont {{Y. Jompol}}, \citenamefont {{I. Farrer}}, \citenamefont {{D.
  A. Ritchie}},\ and\ \citenamefont {{D. C. Glattli}}}]{roche_harvesting_2015}%
  \BibitemOpen
  \bibfield  {author} {\bibinfo {author} {\bibnamefont {{B. Roche}}}, \bibinfo
  {author} {\bibnamefont {{P. Roulleau}}}, \bibinfo {author} {\bibnamefont {{T.
  Jullien}}}, \bibinfo {author} {\bibnamefont {{Y. Jompol}}}, \bibinfo {author}
  {\bibnamefont {{I. Farrer}}}, \bibinfo {author} {\bibnamefont {{D. A.
  Ritchie}}}, \ and\ \bibinfo {author} {\bibnamefont {{D. C. Glattli}}},\
  }\href {\doibase 10.1038/ncomms7738} {\bibfield  {journal} {\bibinfo
  {journal} {Nature Communications}\ }\textbf {\bibinfo {volume} {6}},\
  \bibinfo {pages} {6738} (\bibinfo {year} {2015})}\BibitemShut {NoStop}%
\bibitem [{\citenamefont {Humphrey}\ \emph {et~al.}(2002)\citenamefont
  {Humphrey}, \citenamefont {Newbury}, \citenamefont {Taylor},\ and\
  \citenamefont {Linke}}]{humphrey_reversible_2002}%
  \BibitemOpen
  \bibfield  {author} {\bibinfo {author} {\bibfnamefont {T.~E.}\ \bibnamefont
  {Humphrey}}, \bibinfo {author} {\bibfnamefont {R.}~\bibnamefont {Newbury}},
  \bibinfo {author} {\bibfnamefont {R.~P.}\ \bibnamefont {Taylor}}, \ and\
  \bibinfo {author} {\bibfnamefont {H.}~\bibnamefont {Linke}},\ }\href
  {\doibase 10.1103/PhysRevLett.89.116801} {\bibfield  {journal} {\bibinfo
  {journal} {Phys. Rev. Lett.}\ }\textbf {\bibinfo {volume} {89}},\ \bibinfo
  {pages} {116801} (\bibinfo {year} {2002})}\BibitemShut {NoStop}%
\bibitem [{\citenamefont {Josefsson}\ \emph {et~al.}(2018)\citenamefont
  {Josefsson}, \citenamefont {Svilans}, \citenamefont {Burke}, \citenamefont
  {Hoffmann}, \citenamefont {Fahlvik}, \citenamefont {Thelander}, \citenamefont
  {Leijnse},\ and\ \citenamefont {Linke}}]{josefsson_quantum-dot_2018}%
  \BibitemOpen
  \bibfield  {author} {\bibinfo {author} {\bibfnamefont {M.}~\bibnamefont
  {Josefsson}}, \bibinfo {author} {\bibfnamefont {A.}~\bibnamefont {Svilans}},
  \bibinfo {author} {\bibfnamefont {A.~M.}\ \bibnamefont {Burke}}, \bibinfo
  {author} {\bibfnamefont {E.~A.}\ \bibnamefont {Hoffmann}}, \bibinfo {author}
  {\bibfnamefont {S.}~\bibnamefont {Fahlvik}}, \bibinfo {author} {\bibfnamefont
  {C.}~\bibnamefont {Thelander}}, \bibinfo {author} {\bibfnamefont
  {M.}~\bibnamefont {Leijnse}}, \ and\ \bibinfo {author} {\bibfnamefont
  {H.}~\bibnamefont {Linke}},\ }\href {\doibase 10.1038/s41565-018-0200-5}
  {\bibfield  {journal} {\bibinfo  {journal} {Nat. Nanotechnol.}\ }\textbf
  {\bibinfo {volume} {13}},\ \bibinfo {pages} {920} (\bibinfo {year}
  {2018})}\BibitemShut {NoStop}%
\bibitem [{\citenamefont {Jordan}\ \emph {et~al.}(2013)\citenamefont {Jordan},
  \citenamefont {Sothmann}, \citenamefont
  {S{\ifmmode\acute{a}\else\'{a}\fi}nchez},\ and\ \citenamefont
  {B{\ifmmode\ddot{u}\else\"{u}\fi}ttiker}}]{resonant}%
  \BibitemOpen
  \bibfield  {author} {\bibinfo {author} {\bibfnamefont {A.~N.}\ \bibnamefont
  {Jordan}}, \bibinfo {author} {\bibfnamefont {B.}~\bibnamefont {Sothmann}},
  \bibinfo {author} {\bibfnamefont {R.}~\bibnamefont
  {S{\ifmmode\acute{a}\else\'{a}\fi}nchez}}, \ and\ \bibinfo {author}
  {\bibfnamefont {M.}~\bibnamefont {B{\ifmmode\ddot{u}\else\"{u}\fi}ttiker}},\
  }\href {\doibase 10.1103/PhysRevB.87.075312} {\bibfield  {journal} {\bibinfo
  {journal} {Phys. Rev. B}\ }\textbf {\bibinfo {volume} {87}},\ \bibinfo
  {pages} {075312} (\bibinfo {year} {2013})}\BibitemShut {NoStop}%
\bibitem [{\citenamefont {Jaliel}\ \emph {et~al.}(2019)\citenamefont {Jaliel},
  \citenamefont {Puddy}, \citenamefont
  {S{\ifmmode\acute{a}\else\'{a}\fi}nchez}, \citenamefont {Jordan},
  \citenamefont {Sothmann}, \citenamefont {Farrer}, \citenamefont {Griffiths},
  \citenamefont {Ritchie},\ and\ \citenamefont {Smith}}]{gulzat}%
  \BibitemOpen
  \bibfield  {author} {\bibinfo {author} {\bibfnamefont {G.}~\bibnamefont
  {Jaliel}}, \bibinfo {author} {\bibfnamefont {R.~K.}\ \bibnamefont {Puddy}},
  \bibinfo {author} {\bibfnamefont {R.}~\bibnamefont
  {S{\ifmmode\acute{a}\else\'{a}\fi}nchez}}, \bibinfo {author} {\bibfnamefont
  {A.~N.}\ \bibnamefont {Jordan}}, \bibinfo {author} {\bibfnamefont
  {B.}~\bibnamefont {Sothmann}}, \bibinfo {author} {\bibfnamefont
  {I.}~\bibnamefont {Farrer}}, \bibinfo {author} {\bibfnamefont {J.~P.}\
  \bibnamefont {Griffiths}}, \bibinfo {author} {\bibfnamefont {D.~A.}\
  \bibnamefont {Ritchie}}, \ and\ \bibinfo {author} {\bibfnamefont {C.~G.}\
  \bibnamefont {Smith}},\ }\href {https://arxiv.org/abs/1901.10561} {\bibfield
  {journal} {\bibinfo  {journal} {arXiv}\ } (\bibinfo {year} {2019})},\ \Eprint
  {http://arxiv.org/abs/1901.10561} {1901.10561} \BibitemShut {NoStop}%
\bibitem [{\citenamefont {Courtois}\ \emph {et~al.}(2014)\citenamefont
  {Courtois}, \citenamefont {Hekking}, \citenamefont {Nguyen},\ and\
  \citenamefont {Winkelmann}}]{courtois_electronic_2014}%
  \BibitemOpen
  \bibfield  {author} {\bibinfo {author} {\bibfnamefont {H.}~\bibnamefont
  {Courtois}}, \bibinfo {author} {\bibfnamefont {F.~W.~J.}\ \bibnamefont
  {Hekking}}, \bibinfo {author} {\bibfnamefont {H.~Q.}\ \bibnamefont {Nguyen}},
  \ and\ \bibinfo {author} {\bibfnamefont {C.~B.}\ \bibnamefont {Winkelmann}},\
  }\href {\doibase 10.1007/s10909-014-1101-0} {\bibfield  {journal} {\bibinfo
  {journal} {J. Low Temp. Phys.}\ }\textbf {\bibinfo {volume} {175}},\ \bibinfo
  {pages} {799} (\bibinfo {year} {2014})}\BibitemShut {NoStop}%
\bibitem [{\citenamefont {Edwards}\ \emph {et~al.}(1993)\citenamefont
  {Edwards}, \citenamefont {Niu},\ and\ \citenamefont
  {de~Lozanne}}]{edwards_quantum_1993}%
  \BibitemOpen
  \bibfield  {author} {\bibinfo {author} {\bibfnamefont {H.~L.}\ \bibnamefont
  {Edwards}}, \bibinfo {author} {\bibfnamefont {Q.}~\bibnamefont {Niu}}, \ and\
  \bibinfo {author} {\bibfnamefont {A.~L.}\ \bibnamefont {de~Lozanne}},\ }\href
  {\doibase 10.1063/1.110672} {\bibfield  {journal} {\bibinfo  {journal} {Appl.
  Phys. Lett.}\ }\textbf {\bibinfo {volume} {63}},\ \bibinfo {pages} {1815}
  (\bibinfo {year} {1993})}\BibitemShut {NoStop}%
\bibitem [{\citenamefont {Prance}\ \emph {et~al.}(2009)\citenamefont {Prance},
  \citenamefont {Smith}, \citenamefont {Griffiths}, \citenamefont {Chorley},
  \citenamefont {Anderson}, \citenamefont {Jones}, \citenamefont {Farrer},\
  and\ \citenamefont {Ritchie}}]{prance_electronic_2009}%
  \BibitemOpen
  \bibfield  {author} {\bibinfo {author} {\bibfnamefont {J.~R.}\ \bibnamefont
  {Prance}}, \bibinfo {author} {\bibfnamefont {C.~G.}\ \bibnamefont {Smith}},
  \bibinfo {author} {\bibfnamefont {J.~P.}\ \bibnamefont {Griffiths}}, \bibinfo
  {author} {\bibfnamefont {S.~J.}\ \bibnamefont {Chorley}}, \bibinfo {author}
  {\bibfnamefont {D.}~\bibnamefont {Anderson}}, \bibinfo {author}
  {\bibfnamefont {G.~A.~C.}\ \bibnamefont {Jones}}, \bibinfo {author}
  {\bibfnamefont {I.}~\bibnamefont {Farrer}}, \ and\ \bibinfo {author}
  {\bibfnamefont {D.~A.}\ \bibnamefont {Ritchie}},\ }\href {\doibase
  10.1103/PhysRevLett.102.146602} {\bibfield  {journal} {\bibinfo  {journal}
  {Phys. Rev. Lett.}\ }\textbf {\bibinfo {volume} {102}},\ \bibinfo {pages}
  {146602} (\bibinfo {year} {2009})}\BibitemShut {NoStop}%
\bibitem [{\citenamefont {Nahum}\ \emph {et~al.}(1994)\citenamefont {Nahum},
  \citenamefont {Eiles},\ and\ \citenamefont
  {Martinis}}]{nahum_electronic_1994}%
  \BibitemOpen
  \bibfield  {author} {\bibinfo {author} {\bibfnamefont {M.}~\bibnamefont
  {Nahum}}, \bibinfo {author} {\bibfnamefont {T.~M.}\ \bibnamefont {Eiles}}, \
  and\ \bibinfo {author} {\bibfnamefont {J.~M.}\ \bibnamefont {Martinis}},\
  }\href {\doibase 10.1063/1.112456} {\bibfield  {journal} {\bibinfo  {journal}
  {Appl. Phys. Lett.}\ }\textbf {\bibinfo {volume} {65}},\ \bibinfo {pages}
  {3123} (\bibinfo {year} {1994})}\BibitemShut {NoStop}%
\bibitem [{\citenamefont {Leivo}\ \emph {et~al.}(1996)\citenamefont {Leivo},
  \citenamefont {Pekola},\ and\ \citenamefont {Averin}}]{leivo_efficient_1996}%
  \BibitemOpen
  \bibfield  {author} {\bibinfo {author} {\bibfnamefont {M.~M.}\ \bibnamefont
  {Leivo}}, \bibinfo {author} {\bibfnamefont {J.~P.}\ \bibnamefont {Pekola}}, \
  and\ \bibinfo {author} {\bibfnamefont {D.~V.}\ \bibnamefont {Averin}},\
  }\href {\doibase 10.1063/1.115651} {\bibfield  {journal} {\bibinfo  {journal}
  {Appl. Phys. Lett.}\ }\textbf {\bibinfo {volume} {68}},\ \bibinfo {pages}
  {1996} (\bibinfo {year} {1996})}\BibitemShut {NoStop}%
\bibitem [{\citenamefont {Saira}\ \emph {et~al.}(2007)\citenamefont {Saira},
  \citenamefont {Meschke}, \citenamefont {Giazotto}, \citenamefont {Savin},
  \citenamefont
  {M{\ifmmode\ddot{o}\else\"{o}\fi}tt{\ifmmode\ddot{o}\else\"{o}\fi}nen},\ and\
  \citenamefont {Pekola}}]{saira_heat_2007}%
  \BibitemOpen
  \bibfield  {author} {\bibinfo {author} {\bibfnamefont {O.-P.}\ \bibnamefont
  {Saira}}, \bibinfo {author} {\bibfnamefont {M.}~\bibnamefont {Meschke}},
  \bibinfo {author} {\bibfnamefont {F.}~\bibnamefont {Giazotto}}, \bibinfo
  {author} {\bibfnamefont {A.~M.}\ \bibnamefont {Savin}}, \bibinfo {author}
  {\bibfnamefont {M.}~\bibnamefont
  {M{\ifmmode\ddot{o}\else\"{o}\fi}tt{\ifmmode\ddot{o}\else\"{o}\fi}nen}}, \
  and\ \bibinfo {author} {\bibfnamefont {J.~P.}\ \bibnamefont {Pekola}},\
  }\href {\doibase 10.1103/PhysRevLett.99.027203} {\bibfield  {journal}
  {\bibinfo  {journal} {Phys. Rev. Lett.}\ }\textbf {\bibinfo {volume} {99}},\
  \bibinfo {pages} {027203} (\bibinfo {year} {2007})}\BibitemShut {NoStop}%
\bibitem [{\citenamefont {Pekola}\ \emph {et~al.}(2014)\citenamefont {Pekola},
  \citenamefont {Koski},\ and\ \citenamefont
  {Averin}}]{pekola_refrigerator_2014}%
  \BibitemOpen
  \bibfield  {author} {\bibinfo {author} {\bibfnamefont {J.~P.}\ \bibnamefont
  {Pekola}}, \bibinfo {author} {\bibfnamefont {J.~V.}\ \bibnamefont {Koski}}, \
  and\ \bibinfo {author} {\bibfnamefont {D.~V.}\ \bibnamefont {Averin}},\
  }\href {\doibase 10.1103/PhysRevB.89.081309} {\bibfield  {journal} {\bibinfo
  {journal} {Phys. Rev. B}\ }\textbf {\bibinfo {volume} {89}},\ \bibinfo
  {pages} {081309} (\bibinfo {year} {2014})}\BibitemShut {NoStop}%
\bibitem [{\citenamefont {Feshchenko}\ \emph {et~al.}(2014)\citenamefont
  {Feshchenko}, \citenamefont {Koski},\ and\ \citenamefont
  {Pekola}}]{feshchenko_experimental_2014}%
  \BibitemOpen
  \bibfield  {author} {\bibinfo {author} {\bibfnamefont {A.~V.}\ \bibnamefont
  {Feshchenko}}, \bibinfo {author} {\bibfnamefont {J.~V.}\ \bibnamefont
  {Koski}}, \ and\ \bibinfo {author} {\bibfnamefont {J.~P.}\ \bibnamefont
  {Pekola}},\ }\href {\doibase 10.1103/PhysRevB.90.201407} {\bibfield
  {journal} {\bibinfo  {journal} {Phys. Rev. B}\ }\textbf {\bibinfo {volume}
  {90}},\ \bibinfo {pages} {201407} (\bibinfo {year} {2014})}\BibitemShut
  {NoStop}%
\bibitem [{\citenamefont {Koski}\ \emph {et~al.}(2015)\citenamefont {Koski},
  \citenamefont {Kutvonen}, \citenamefont {Khaymovich}, \citenamefont
  {Ala-Nissila},\ and\ \citenamefont {Pekola}}]{koski_onchip_2015}%
  \BibitemOpen
  \bibfield  {author} {\bibinfo {author} {\bibfnamefont {J.~V.}\ \bibnamefont
  {Koski}}, \bibinfo {author} {\bibfnamefont {A.}~\bibnamefont {Kutvonen}},
  \bibinfo {author} {\bibfnamefont {I.~M.}\ \bibnamefont {Khaymovich}},
  \bibinfo {author} {\bibfnamefont {T.}~\bibnamefont {Ala-Nissila}}, \ and\
  \bibinfo {author} {\bibfnamefont {J.~P.}\ \bibnamefont {Pekola}},\ }\href
  {\doibase 10.1103/PhysRevLett.115.260602} {\bibfield  {journal} {\bibinfo
  {journal} {Phys. Rev. Lett.}\ }\textbf {\bibinfo {volume} {115}},\ \bibinfo
  {pages} {260602} (\bibinfo {year} {2015})}\BibitemShut {NoStop}%
\bibitem [{\citenamefont {Thierschmann}\ \emph {et~al.}(2015)\citenamefont
  {Thierschmann}, \citenamefont {Arnold}, \citenamefont {Mitterm\"uller},
  \citenamefont {Maier}, \citenamefont {Heyn}, \citenamefont {Hansen},
  \citenamefont {Buhmann},\ and\ \citenamefont
  {Molenkamp}}]{thierschmann_thermal_2015}%
  \BibitemOpen
  \bibfield  {author} {\bibinfo {author} {\bibfnamefont {H.}~\bibnamefont
  {Thierschmann}}, \bibinfo {author} {\bibfnamefont {F.}~\bibnamefont
  {Arnold}}, \bibinfo {author} {\bibfnamefont {M.}~\bibnamefont
  {Mitterm\"uller}}, \bibinfo {author} {\bibfnamefont {L.}~\bibnamefont
  {Maier}}, \bibinfo {author} {\bibfnamefont {C.}~\bibnamefont {Heyn}},
  \bibinfo {author} {\bibfnamefont {W.}~\bibnamefont {Hansen}}, \bibinfo
  {author} {\bibfnamefont {H.}~\bibnamefont {Buhmann}}, \ and\ \bibinfo
  {author} {\bibfnamefont {L.~W.}\ \bibnamefont {Molenkamp}},\ }\href
  {http://stacks.iop.org/1367-2630/17/i=11/a=113003} {\bibfield  {journal}
  {\bibinfo  {journal} {New J. Phys.}\ }\textbf {\bibinfo {volume} {17}},\
  \bibinfo {pages} {113003} (\bibinfo {year} {2015})}\BibitemShut {NoStop}%
\bibitem [{\citenamefont {S\'anchez}\ \emph {et~al.}(2017)\citenamefont
  {S\'anchez}, \citenamefont {Thierschmann},\ and\ \citenamefont
  {Molenkamp}}]{transistor}%
  \BibitemOpen
  \bibfield  {author} {\bibinfo {author} {\bibfnamefont {R.}~\bibnamefont
  {S\'anchez}}, \bibinfo {author} {\bibfnamefont {H.}~\bibnamefont
  {Thierschmann}}, \ and\ \bibinfo {author} {\bibfnamefont {L.~W.}\
  \bibnamefont {Molenkamp}},\ }\href {\doibase 10.1103/PhysRevB.95.241401}
  {\bibfield  {journal} {\bibinfo  {journal} {Phys. Rev. B}\ }\textbf {\bibinfo
  {volume} {95}},\ \bibinfo {pages} {241401} (\bibinfo {year}
  {2017})}\BibitemShut {NoStop}%
\bibitem [{\citenamefont {Scheibner}\ \emph {et~al.}(2008)\citenamefont
  {Scheibner}, \citenamefont {K{\ifmmode\ddot{o}\else\"{o}\fi}nig},
  \citenamefont {Reuter}, \citenamefont {Wieck}, \citenamefont {Gould},
  \citenamefont {Buhmann},\ and\ \citenamefont
  {Molenkamp}}]{scheibner_quantum_2008}%
  \BibitemOpen
  \bibfield  {author} {\bibinfo {author} {\bibfnamefont {R.}~\bibnamefont
  {Scheibner}}, \bibinfo {author} {\bibfnamefont {M.}~\bibnamefont
  {K{\ifmmode\ddot{o}\else\"{o}\fi}nig}}, \bibinfo {author} {\bibfnamefont
  {D.}~\bibnamefont {Reuter}}, \bibinfo {author} {\bibfnamefont {A.~D.}\
  \bibnamefont {Wieck}}, \bibinfo {author} {\bibfnamefont {C.}~\bibnamefont
  {Gould}}, \bibinfo {author} {\bibfnamefont {H.}~\bibnamefont {Buhmann}}, \
  and\ \bibinfo {author} {\bibfnamefont {L.~W.}\ \bibnamefont {Molenkamp}},\
  }\href {\doibase 10.1088/1367-2630/10/8/083016} {\bibfield  {journal}
  {\bibinfo  {journal} {New J. Phys.}\ }\textbf {\bibinfo {volume} {10}},\
  \bibinfo {pages} {083016} (\bibinfo {year} {2008})}\BibitemShut {NoStop}%
\bibitem [{\citenamefont {Giazotto}\ and\ \citenamefont
  {Bergeret}(2013)}]{giazotto_thermal_2013}%
  \BibitemOpen
  \bibfield  {author} {\bibinfo {author} {\bibfnamefont {F.}~\bibnamefont
  {Giazotto}}\ and\ \bibinfo {author} {\bibfnamefont {F.~S.}\ \bibnamefont
  {Bergeret}},\ }\href {\doibase 10.1063/1.4846375} {\bibfield  {journal}
  {\bibinfo  {journal} {Appl. Phys. Lett.}\ }\textbf {\bibinfo {volume}
  {103}},\ \bibinfo {pages} {242602} (\bibinfo {year} {2013})}\BibitemShut
  {NoStop}%
\bibitem [{\citenamefont
  {Mart{\ifmmode\acute{\imath}\else\'{\i}\fi}nez-P{\ifmmode\acute{e}\else\'{e}\fi}rez}\
  and\ \citenamefont {Giazotto}(2013)}]{martinez-perez_efficient_2013}%
  \BibitemOpen
  \bibfield  {author} {\bibinfo {author} {\bibfnamefont {M.~J.}\ \bibnamefont
  {Mart{\ifmmode\acute{\imath}\else\'{\i}\fi}nez-P{\ifmmode\acute{e}\else\'{e}\fi}rez}}\
  and\ \bibinfo {author} {\bibfnamefont {F.}~\bibnamefont {Giazotto}},\ }\href
  {\doibase 10.1063/1.4804550} {\bibfield  {journal} {\bibinfo  {journal}
  {Appl. Phys. Lett.}\ }\textbf {\bibinfo {volume} {102}},\ \bibinfo {pages}
  {182602} (\bibinfo {year} {2013})}\BibitemShut {NoStop}%
\bibitem [{\citenamefont
  {Mart{\ifmmode\acute{\imath}\else\'{\i}\fi}nez-P{\ifmmode\acute{e}\else\'{e}\fi}rez}\
  \emph {et~al.}(2015)\citenamefont
  {Mart{\ifmmode\acute{\imath}\else\'{\i}\fi}nez-P{\ifmmode\acute{e}\else\'{e}\fi}rez},
  \citenamefont {Fornieri},\ and\ \citenamefont
  {Giazotto}}]{martinez-perez_rectification_2015}%
  \BibitemOpen
  \bibfield  {author} {\bibinfo {author} {\bibfnamefont {M.~J.}\ \bibnamefont
  {Mart{\ifmmode\acute{\imath}\else\'{\i}\fi}nez-P{\ifmmode\acute{e}\else\'{e}\fi}rez}},
  \bibinfo {author} {\bibfnamefont {A.}~\bibnamefont {Fornieri}}, \ and\
  \bibinfo {author} {\bibfnamefont {F.}~\bibnamefont {Giazotto}},\ }\href
  {\doibase 10.1038/nnano.2015.11} {\bibfield  {journal} {\bibinfo  {journal}
  {Nat. Nanotechnol.}\ }\textbf {\bibinfo {volume} {10}},\ \bibinfo {pages}
  {303} (\bibinfo {year} {2015})}\BibitemShut {NoStop}%
\bibitem [{\citenamefont {Ronzani}\ \emph {et~al.}(2018)\citenamefont
  {Ronzani}, \citenamefont {Karimi}, \citenamefont {Senior}, \citenamefont
  {Chang}, \citenamefont {Peltonen}, \citenamefont {Chen},\ and\ \citenamefont
  {Pekola}}]{ronzani_tunable_2018}%
  \BibitemOpen
  \bibfield  {author} {\bibinfo {author} {\bibfnamefont {A.}~\bibnamefont
  {Ronzani}}, \bibinfo {author} {\bibfnamefont {B.}~\bibnamefont {Karimi}},
  \bibinfo {author} {\bibfnamefont {J.}~\bibnamefont {Senior}}, \bibinfo
  {author} {\bibfnamefont {Y.-C.}\ \bibnamefont {Chang}}, \bibinfo {author}
  {\bibfnamefont {J.~T.}\ \bibnamefont {Peltonen}}, \bibinfo {author}
  {\bibfnamefont {C.}~\bibnamefont {Chen}}, \ and\ \bibinfo {author}
  {\bibfnamefont {J.~P.}\ \bibnamefont {Pekola}},\ }\href {\doibase
  10.1038/s41567-018-0199-4} {\bibfield  {journal} {\bibinfo  {journal} {Nat.
  Phys.}\ }\textbf {\bibinfo {volume} {14}},\ \bibinfo {pages} {991} (\bibinfo
  {year} {2018})}\BibitemShut {NoStop}%
\bibitem [{\citenamefont {Benenti}\ \emph {et~al.}(2016)\citenamefont
  {Benenti}, \citenamefont {Casati}, \citenamefont {Mej{\'i}a-Monasterio},\
  and\ \citenamefont {Peyrard}}]{benenti_from_2016}%
  \BibitemOpen
  \bibfield  {author} {\bibinfo {author} {\bibfnamefont {G.}~\bibnamefont
  {Benenti}}, \bibinfo {author} {\bibfnamefont {G.}~\bibnamefont {Casati}},
  \bibinfo {author} {\bibfnamefont {C.}~\bibnamefont {Mej{\'i}a-Monasterio}}, \
  and\ \bibinfo {author} {\bibfnamefont {M.}~\bibnamefont {Peyrard}},\
  }\enquote {\bibinfo {title} {From thermal rectifiers to thermoelectric
  devices},}\ in\ \href {\doibase 10.1007/978-3-319-29261-8_10} {\emph
  {\bibinfo {booktitle} {Thermal Transport in Low Dimensions: From Statistical
  Physics to Nanoscale Heat Transfer}}},\ \bibinfo {editor} {edited by\
  \bibinfo {editor} {\bibfnamefont {S.}~\bibnamefont {Lepri}}}\ (\bibinfo
  {publisher} {Springer International Publishing},\ \bibinfo {address} {Cham},\
  \bibinfo {year} {2016})\ pp.\ \bibinfo {pages} {365--407}\BibitemShut
  {NoStop}%
\bibitem [{\citenamefont {Ruokola}\ \emph {et~al.}(2009)\citenamefont
  {Ruokola}, \citenamefont {Ojanen},\ and\ \citenamefont
  {Jauho}}]{ruokola_thermal_2009}%
  \BibitemOpen
  \bibfield  {author} {\bibinfo {author} {\bibfnamefont {T.}~\bibnamefont
  {Ruokola}}, \bibinfo {author} {\bibfnamefont {T.}~\bibnamefont {Ojanen}}, \
  and\ \bibinfo {author} {\bibfnamefont {A.-P.}\ \bibnamefont {Jauho}},\ }\href
  {\doibase 10.1103/PhysRevB.79.144306} {\bibfield  {journal} {\bibinfo
  {journal} {Phys. Rev. B}\ }\textbf {\bibinfo {volume} {79}},\ \bibinfo
  {pages} {144306} (\bibinfo {year} {2009})}\BibitemShut {NoStop}%
\bibitem [{\citenamefont {Sierra}\ and\ \citenamefont
  {S{\ifmmode\acute{a}\else\'{a}\fi}nchez}(2015)}]{sierra_nonlinear_2015}%
  \BibitemOpen
  \bibfield  {author} {\bibinfo {author} {\bibfnamefont {M.~A.}\ \bibnamefont
  {Sierra}}\ and\ \bibinfo {author} {\bibfnamefont {D.}~\bibnamefont
  {S{\ifmmode\acute{a}\else\'{a}\fi}nchez}},\ }\href {\doibase
  10.1016/j.matpr.2015.05.066} {\bibfield  {journal} {\bibinfo  {journal}
  {Mater. Today:. Proc.}\ }\textbf {\bibinfo {volume} {2}},\ \bibinfo {pages}
  {483} (\bibinfo {year} {2015})}\BibitemShut {NoStop}%
\bibitem [{\citenamefont {Vannucci}\ \emph {et~al.}(2015)\citenamefont
  {Vannucci}, \citenamefont {Ronetti}, \citenamefont {Dolcetto}, \citenamefont
  {Carrega},\ and\ \citenamefont {Sassetti}}]{vannucci_interference_2015}%
  \BibitemOpen
  \bibfield  {author} {\bibinfo {author} {\bibfnamefont {L.}~\bibnamefont
  {Vannucci}}, \bibinfo {author} {\bibfnamefont {F.}~\bibnamefont {Ronetti}},
  \bibinfo {author} {\bibfnamefont {G.}~\bibnamefont {Dolcetto}}, \bibinfo
  {author} {\bibfnamefont {M.}~\bibnamefont {Carrega}}, \ and\ \bibinfo
  {author} {\bibfnamefont {M.}~\bibnamefont {Sassetti}},\ }\href {\doibase
  10.1103/PhysRevB.92.075446} {\bibfield  {journal} {\bibinfo  {journal} {Phys.
  Rev. B}\ }\textbf {\bibinfo {volume} {92}},\ \bibinfo {pages} {075446}
  (\bibinfo {year} {2015})}\BibitemShut {NoStop}%
\bibitem [{\citenamefont {Marcos-Vicioso}\ \emph {et~al.}(2018)\citenamefont
  {Marcos-Vicioso}, \citenamefont
  {L{\ifmmode\acute{o}\else\'{o}\fi}pez-Jurado}, \citenamefont {Ruiz-Garcia},\
  and\ \citenamefont {S{\ifmmode\acute{a}\else\'{a}\fi}nchez}}]{tfg}%
  \BibitemOpen
  \bibfield  {author} {\bibinfo {author} {\bibfnamefont {A.}~\bibnamefont
  {Marcos-Vicioso}}, \bibinfo {author} {\bibfnamefont {C.}~\bibnamefont
  {L{\ifmmode\acute{o}\else\'{o}\fi}pez-Jurado}}, \bibinfo {author}
  {\bibfnamefont {M.}~\bibnamefont {Ruiz-Garcia}}, \ and\ \bibinfo {author}
  {\bibfnamefont {R.}~\bibnamefont {S{\ifmmode\acute{a}\else\'{a}\fi}nchez}},\
  }\href {\doibase 10.1103/PhysRevB.98.035414} {\bibfield  {journal} {\bibinfo
  {journal} {Phys. Rev. B}\ }\textbf {\bibinfo {volume} {98}},\ \bibinfo
  {pages} {035414} (\bibinfo {year} {2018})}\BibitemShut {NoStop}%
\bibitem [{\citenamefont {Fornieri}\ \emph {et~al.}(2014)\citenamefont
  {Fornieri}, \citenamefont
  {Mart{\ifmmode\acute{\imath}\else\'{\i}\fi}nez-P{\ifmmode\acute{e}\else\'{e}\fi}rez},\
  and\ \citenamefont {Giazotto}}]{fornieri_normal_2014}%
  \BibitemOpen
  \bibfield  {author} {\bibinfo {author} {\bibfnamefont {A.}~\bibnamefont
  {Fornieri}}, \bibinfo {author} {\bibfnamefont {M.~J.}\ \bibnamefont
  {Mart{\ifmmode\acute{\imath}\else\'{\i}\fi}nez-P{\ifmmode\acute{e}\else\'{e}\fi}rez}},
  \ and\ \bibinfo {author} {\bibfnamefont {F.}~\bibnamefont {Giazotto}},\
  }\href {\doibase 10.1063/1.4875917} {\bibfield  {journal} {\bibinfo
  {journal} {Appl. Phys. Lett.}\ }\textbf {\bibinfo {volume} {104}},\ \bibinfo
  {pages} {183108} (\bibinfo {year} {2014})}\BibitemShut {NoStop}%
\bibitem [{\citenamefont {S{\ifmmode\acute{a}\else\'{a}\fi}nchez}\ \emph
  {et~al.}(2015)\citenamefont {S{\ifmmode\acute{a}\else\'{a}\fi}nchez},
  \citenamefont {Sothmann},\ and\ \citenamefont {Jordan}}]{diode}%
  \BibitemOpen
  \bibfield  {author} {\bibinfo {author} {\bibfnamefont {R.}~\bibnamefont
  {S{\ifmmode\acute{a}\else\'{a}\fi}nchez}}, \bibinfo {author} {\bibfnamefont
  {B.}~\bibnamefont {Sothmann}}, \ and\ \bibinfo {author} {\bibfnamefont
  {A.~N.}\ \bibnamefont {Jordan}},\ }\href {\doibase
  10.1088/1367-2630/17/7/075006} {\bibfield  {journal} {\bibinfo  {journal}
  {New J. Phys.}\ }\textbf {\bibinfo {volume} {17}},\ \bibinfo {pages} {075006}
  (\bibinfo {year} {2015})}\BibitemShut {NoStop}%
\bibitem [{\citenamefont {Jiang}\ \emph {et~al.}(2015)\citenamefont {Jiang},
  \citenamefont {Kulkarni}, \citenamefont {Segal},\ and\ \citenamefont
  {Imry}}]{jiang_phonon_2015}%
  \BibitemOpen
  \bibfield  {author} {\bibinfo {author} {\bibfnamefont {J.-H.}\ \bibnamefont
  {Jiang}}, \bibinfo {author} {\bibfnamefont {M.}~\bibnamefont {Kulkarni}},
  \bibinfo {author} {\bibfnamefont {D.}~\bibnamefont {Segal}}, \ and\ \bibinfo
  {author} {\bibfnamefont {Y.}~\bibnamefont {Imry}},\ }\href {\doibase
  10.1103/PhysRevB.92.045309} {\bibfield  {journal} {\bibinfo  {journal} {Phys.
  Rev. B}\ }\textbf {\bibinfo {volume} {92}},\ \bibinfo {pages} {045309}
  (\bibinfo {year} {2015})}\BibitemShut {NoStop}%
\bibitem [{\citenamefont {Rossell{\ifmmode\acute{o}\else\'{o}\fi}}\ \emph
  {et~al.}(2017)\citenamefont {Rossell{\ifmmode\acute{o}\else\'{o}\fi}},
  \citenamefont {L{\ifmmode\acute{o}\else\'{o}\fi}pez},\ and\ \citenamefont
  {S{\ifmmode\acute{a}\else\'{a}\fi}nchez}}]{guillem}%
  \BibitemOpen
  \bibfield  {author} {\bibinfo {author} {\bibfnamefont {G.}~\bibnamefont
  {Rossell{\ifmmode\acute{o}\else\'{o}\fi}}}, \bibinfo {author} {\bibfnamefont
  {R.}~\bibnamefont {L{\ifmmode\acute{o}\else\'{o}\fi}pez}}, \ and\ \bibinfo
  {author} {\bibfnamefont {R.}~\bibnamefont
  {S{\ifmmode\acute{a}\else\'{a}\fi}nchez}},\ }\href {\doibase
  10.1103/PhysRevB.95.235404} {\bibfield  {journal} {\bibinfo  {journal} {Phys.
  Rev. B}\ }\textbf {\bibinfo {volume} {95}},\ \bibinfo {pages} {235404}
  (\bibinfo {year} {2017})}\BibitemShut {NoStop}%
\bibitem [{\citenamefont {S{\ifmmode\acute{a}\else\'{a}\fi}nchez}\ \emph
  {et~al.}(2017)\citenamefont {S{\ifmmode\acute{a}\else\'{a}\fi}nchez},
  \citenamefont {Thierschmann},\ and\ \citenamefont {Molenkamp}}]{devices}%
  \BibitemOpen
  \bibfield  {author} {\bibinfo {author} {\bibfnamefont {R.}~\bibnamefont
  {S{\ifmmode\acute{a}\else\'{a}\fi}nchez}}, \bibinfo {author} {\bibfnamefont
  {H.}~\bibnamefont {Thierschmann}}, \ and\ \bibinfo {author} {\bibfnamefont
  {L.~W.}\ \bibnamefont {Molenkamp}},\ }\href {\doibase
  10.1088/1367-2630/aa8b94} {\bibfield  {journal} {\bibinfo  {journal} {New J.
  Phys.}\ }\textbf {\bibinfo {volume} {19}},\ \bibinfo {pages} {113040}
  (\bibinfo {year} {2017})}\BibitemShut {NoStop}%
\bibitem [{\citenamefont {Oettinger}\ \emph {et~al.}(2014)\citenamefont
  {Oettinger}, \citenamefont {Chitra},\ and\ \citenamefont
  {Restrepo}}]{oettinger_heat_2014}%
  \BibitemOpen
  \bibfield  {author} {\bibinfo {author} {\bibfnamefont {D.}~\bibnamefont
  {Oettinger}}, \bibinfo {author} {\bibfnamefont {R.}~\bibnamefont {Chitra}}, \
  and\ \bibinfo {author} {\bibfnamefont {J.}~\bibnamefont {Restrepo}},\ }\href
  {\doibase 10.1140/epjb/e2014-50310-3} {\bibfield  {journal} {\bibinfo
  {journal} {Eur. Phys. J. B}\ }\textbf {\bibinfo {volume} {87}},\ \bibinfo
  {pages} {224} (\bibinfo {year} {2014})}\BibitemShut {NoStop}%
\bibitem [{\citenamefont {Bours}\ \emph {et~al.}(2019)\citenamefont {Bours},
  \citenamefont {Sothmann}, \citenamefont {Carrega}, \citenamefont {Strambini},
  \citenamefont {Braggio}, \citenamefont {Hankiewicz}, \citenamefont
  {Molenkamp},\ and\ \citenamefont {Giazotto}}]{bours_phase-tunable2019}%
  \BibitemOpen
  \bibfield  {author} {\bibinfo {author} {\bibfnamefont {L.}~\bibnamefont
  {Bours}}, \bibinfo {author} {\bibfnamefont {B.}~\bibnamefont {Sothmann}},
  \bibinfo {author} {\bibfnamefont {M.}~\bibnamefont {Carrega}}, \bibinfo
  {author} {\bibfnamefont {E.}~\bibnamefont {Strambini}}, \bibinfo {author}
  {\bibfnamefont {A.}~\bibnamefont {Braggio}}, \bibinfo {author} {\bibfnamefont
  {E.~M.}\ \bibnamefont {Hankiewicz}}, \bibinfo {author} {\bibfnamefont
  {L.~W.}\ \bibnamefont {Molenkamp}}, \ and\ \bibinfo {author} {\bibfnamefont
  {F.}~\bibnamefont {Giazotto}},\ }\href {\doibase
  10.1103/PhysRevApplied.11.044073} {\bibfield  {journal} {\bibinfo  {journal}
  {Phys. Rev. Appl.}\ }\textbf {\bibinfo {volume} {11}},\ \bibinfo {pages}
  {044073} (\bibinfo {year} {2019})}\BibitemShut {NoStop}%
\bibitem [{\citenamefont {Averin}\ and\ \citenamefont
  {Likharev}(1986)}]{averin_coulomb_1986}%
  \BibitemOpen
  \bibfield  {author} {\bibinfo {author} {\bibfnamefont {D.~V.}\ \bibnamefont
  {Averin}}\ and\ \bibinfo {author} {\bibfnamefont {K.~K.}\ \bibnamefont
  {Likharev}},\ }\href {\doibase 10.1007/BF00683469} {\bibfield  {journal}
  {\bibinfo  {journal} {J. Low Temp. Phys.}\ }\textbf {\bibinfo {volume}
  {62}},\ \bibinfo {pages} {345} (\bibinfo {year} {1986})}\BibitemShut
  {NoStop}%
\bibitem [{\citenamefont {Enrico}\ and\ \citenamefont
  {Giazotto}(2016)}]{enrico_superconducting_2016}%
  \BibitemOpen
  \bibfield  {author} {\bibinfo {author} {\bibfnamefont {E.}~\bibnamefont
  {Enrico}}\ and\ \bibinfo {author} {\bibfnamefont {F.}~\bibnamefont
  {Giazotto}},\ }\href {\doibase 10.1103/PhysRevApplied.5.064020} {\bibfield
  {journal} {\bibinfo  {journal} {Phys. Rev. Appl.}\ }\textbf {\bibinfo
  {volume} {5}},\ \bibinfo {pages} {064020} (\bibinfo {year}
  {2016})}\BibitemShut {NoStop}%
\bibitem [{\citenamefont {Enrico}\ \emph {et~al.}(2017)\citenamefont {Enrico},
  \citenamefont {Strambini},\ and\ \citenamefont
  {Giazotto}}]{enrico_phase_2017}%
  \BibitemOpen
  \bibfield  {author} {\bibinfo {author} {\bibfnamefont {E.}~\bibnamefont
  {Enrico}}, \bibinfo {author} {\bibfnamefont {E.}~\bibnamefont {Strambini}}, \
  and\ \bibinfo {author} {\bibfnamefont {F.}~\bibnamefont {Giazotto}},\ }\href
  {\doibase 10.1038/s41598-017-13894-z} {\bibfield  {journal} {\bibinfo
  {journal} {Sci. Rep.}\ }\textbf {\bibinfo {volume} {7}},\ \bibinfo {pages}
  {13492} (\bibinfo {year} {2017})}\BibitemShut {NoStop}%
\bibitem [{\citenamefont {Strambini}\ \emph {et~al.}(2014)\citenamefont
  {Strambini}, \citenamefont {Bergeret},\ and\ \citenamefont
  {Giazotto}}]{strambini_proximity_2014}%
  \BibitemOpen
  \bibfield  {author} {\bibinfo {author} {\bibfnamefont {E.}~\bibnamefont
  {Strambini}}, \bibinfo {author} {\bibfnamefont {F.~S.}\ \bibnamefont
  {Bergeret}}, \ and\ \bibinfo {author} {\bibfnamefont {F.}~\bibnamefont
  {Giazotto}},\ }\href {\doibase 10.1063/1.4893759} {\bibfield  {journal}
  {\bibinfo  {journal} {Appl. Phys. Lett.}\ }\textbf {\bibinfo {volume}
  {105}},\ \bibinfo {pages} {082601} (\bibinfo {year} {2014})}\BibitemShut
  {NoStop}%
\bibitem [{\citenamefont {Heikkil{\ifmmode\ddot{a}\else\"{a}\fi}}\ \emph
  {et~al.}(2002)\citenamefont {Heikkil{\ifmmode\ddot{a}\else\"{a}\fi}},
  \citenamefont
  {S{\ifmmode\ddot{a}\else\"{a}\fi}rkk{\ifmmode\ddot{a}\else\"{a}\fi}},\ and\
  \citenamefont {Wilhelm}}]{heikkila_supercurrent_2002}%
  \BibitemOpen
  \bibfield  {author} {\bibinfo {author} {\bibfnamefont {T.~T.}\ \bibnamefont
  {Heikkil{\ifmmode\ddot{a}\else\"{a}\fi}}}, \bibinfo {author} {\bibfnamefont
  {J.}~\bibnamefont
  {S{\ifmmode\ddot{a}\else\"{a}\fi}rkk{\ifmmode\ddot{a}\else\"{a}\fi}}}, \ and\
  \bibinfo {author} {\bibfnamefont {F.~K.}\ \bibnamefont {Wilhelm}},\ }\href
  {\doibase 10.1103/PhysRevB.66.184513} {\bibfield  {journal} {\bibinfo
  {journal} {Phys. Rev. B}\ }\textbf {\bibinfo {volume} {66}},\ \bibinfo
  {pages} {184513} (\bibinfo {year} {2002})}\BibitemShut {NoStop}%
\bibitem [{\citenamefont {Le~Sueur}\ \emph {et~al.}(2008)\citenamefont
  {Le~Sueur}, \citenamefont {Joyez}, \citenamefont {Pothier}, \citenamefont
  {Urbina},\ and\ \citenamefont {Esteve}}]{leSueur_phase_2008}%
  \BibitemOpen
  \bibfield  {author} {\bibinfo {author} {\bibfnamefont {H.}~\bibnamefont
  {Le~Sueur}}, \bibinfo {author} {\bibfnamefont {P.}~\bibnamefont {Joyez}},
  \bibinfo {author} {\bibfnamefont {H.}~\bibnamefont {Pothier}}, \bibinfo
  {author} {\bibfnamefont {C.}~\bibnamefont {Urbina}}, \ and\ \bibinfo {author}
  {\bibfnamefont {D.}~\bibnamefont {Esteve}},\ }\href {\doibase
  10.1103/PhysRevLett.100.197002} {\bibfield  {journal} {\bibinfo  {journal}
  {Phys. Rev. Lett.}\ }\textbf {\bibinfo {volume} {100}},\ \bibinfo {pages}
  {197002} (\bibinfo {year} {2008})}\BibitemShut {NoStop}%
\bibitem [{\citenamefont {Wellstood}\ \emph {et~al.}(1994)\citenamefont
  {Wellstood}, \citenamefont {Urbina},\ and\ \citenamefont
  {Clarke}}]{wellstood_hot_1994}%
  \BibitemOpen
  \bibfield  {author} {\bibinfo {author} {\bibfnamefont {F.~C.}\ \bibnamefont
  {Wellstood}}, \bibinfo {author} {\bibfnamefont {C.}~\bibnamefont {Urbina}}, \
  and\ \bibinfo {author} {\bibfnamefont {J.}~\bibnamefont {Clarke}},\ }\href
  {\doibase 10.1103/PhysRevB.49.5942} {\bibfield  {journal} {\bibinfo
  {journal} {Phys. Rev. B}\ }\textbf {\bibinfo {volume} {49}},\ \bibinfo
  {pages} {5942} (\bibinfo {year} {1994})}\BibitemShut {NoStop}%
\bibitem [{\citenamefont {Sothmann}\ \emph {et~al.}(2015)\citenamefont
  {Sothmann}, \citenamefont {S\'anchez},\ and\ \citenamefont
  {Jordan}}]{bjorn_review}%
  \BibitemOpen
  \bibfield  {author} {\bibinfo {author} {\bibfnamefont {B.}~\bibnamefont
  {Sothmann}}, \bibinfo {author} {\bibfnamefont {R.}~\bibnamefont {S\'anchez}},
  \ and\ \bibinfo {author} {\bibfnamefont {A.~N.}\ \bibnamefont {Jordan}},\
  }\href {http://stacks.iop.org/0957-4484/26/i=3/a=032001} {\bibfield
  {journal} {\bibinfo  {journal} {Nanotechnology}\ }\textbf {\bibinfo {volume}
  {26}},\ \bibinfo {pages} {032001} (\bibinfo {year} {2015})}\BibitemShut
  {NoStop}%
\bibitem [{fre()}]{freq}%
  \BibitemOpen
  \href@noop {} {}\bibinfo {note} {For typical experimental resistances of
  10--100~k$\Omega$, the tunneling rates set an upper bound on the device
  frequency of around 10--100~GHz.}\BibitemShut {Stop}%
\bibitem [{\citenamefont {Pekola}\ \emph {et~al.}(2010)\citenamefont {Pekola},
  \citenamefont {Maisi}, \citenamefont {Kafanov}, \citenamefont {Chekurov},
  \citenamefont {Kemppinen}, \citenamefont {Pashkin}, \citenamefont {Saira},
  \citenamefont {M\"ott\"onen},\ and\ \citenamefont
  {Tsai}}]{pekola_environment_2010}%
  \BibitemOpen
  \bibfield  {author} {\bibinfo {author} {\bibfnamefont {J.~P.}\ \bibnamefont
  {Pekola}}, \bibinfo {author} {\bibfnamefont {V.~F.}\ \bibnamefont {Maisi}},
  \bibinfo {author} {\bibfnamefont {S.}~\bibnamefont {Kafanov}}, \bibinfo
  {author} {\bibfnamefont {N.}~\bibnamefont {Chekurov}}, \bibinfo {author}
  {\bibfnamefont {A.}~\bibnamefont {Kemppinen}}, \bibinfo {author}
  {\bibfnamefont {Y.~A.}\ \bibnamefont {Pashkin}}, \bibinfo {author}
  {\bibfnamefont {O.-P.}\ \bibnamefont {Saira}}, \bibinfo {author}
  {\bibfnamefont {M.}~\bibnamefont {M\"ott\"onen}}, \ and\ \bibinfo {author}
  {\bibfnamefont {J.~S.}\ \bibnamefont {Tsai}},\ }\href {\doibase
  10.1103/PhysRevLett.105.026803} {\bibfield  {journal} {\bibinfo  {journal}
  {Phys. Rev. Lett.}\ }\textbf {\bibinfo {volume} {105}},\ \bibinfo {pages}
  {026803} (\bibinfo {year} {2010})}\BibitemShut {NoStop}%
\bibitem [{\citenamefont {Timofeev}\ \emph {et~al.}(2009)\citenamefont
  {Timofeev}, \citenamefont {Garc\'{\i}a}, \citenamefont {Kopnin},
  \citenamefont {Savin}, \citenamefont {Meschke}, \citenamefont {Giazotto},\
  and\ \citenamefont {Pekola}}]{timofeev_recombination_2009}%
  \BibitemOpen
  \bibfield  {author} {\bibinfo {author} {\bibfnamefont {A.~V.}\ \bibnamefont
  {Timofeev}}, \bibinfo {author} {\bibfnamefont {C.~P.}\ \bibnamefont
  {Garc\'{\i}a}}, \bibinfo {author} {\bibfnamefont {N.~B.}\ \bibnamefont
  {Kopnin}}, \bibinfo {author} {\bibfnamefont {A.~M.}\ \bibnamefont {Savin}},
  \bibinfo {author} {\bibfnamefont {M.}~\bibnamefont {Meschke}}, \bibinfo
  {author} {\bibfnamefont {F.}~\bibnamefont {Giazotto}}, \ and\ \bibinfo
  {author} {\bibfnamefont {J.~P.}\ \bibnamefont {Pekola}},\ }\href {\doibase
  10.1103/PhysRevLett.102.017003} {\bibfield  {journal} {\bibinfo  {journal}
  {Phys. Rev. Lett.}\ }\textbf {\bibinfo {volume} {102}},\ \bibinfo {pages}
  {017003} (\bibinfo {year} {2009})}\BibitemShut {NoStop}%
\bibitem [{\citenamefont {Staring}\ \emph {et~al.}(1993)\citenamefont
  {Staring}, \citenamefont {Molenkamp}, \citenamefont {Alphenaar},
  \citenamefont {van Houten}, \citenamefont {Buyk}, \citenamefont {Mabesoone},
  \citenamefont {Beenakker},\ and\ \citenamefont
  {Foxon}}]{staring_coulomb-blockade_1993}%
  \BibitemOpen
  \bibfield  {author} {\bibinfo {author} {\bibfnamefont {A.~A.~M.}\
  \bibnamefont {Staring}}, \bibinfo {author} {\bibfnamefont {L.~W.}\
  \bibnamefont {Molenkamp}}, \bibinfo {author} {\bibfnamefont {B.~W.}\
  \bibnamefont {Alphenaar}}, \bibinfo {author} {\bibfnamefont {H.}~\bibnamefont
  {van Houten}}, \bibinfo {author} {\bibfnamefont {O.~J.~A.}\ \bibnamefont
  {Buyk}}, \bibinfo {author} {\bibfnamefont {M.~A.~A.}\ \bibnamefont
  {Mabesoone}}, \bibinfo {author} {\bibfnamefont {C.~W.~J.}\ \bibnamefont
  {Beenakker}}, \ and\ \bibinfo {author} {\bibfnamefont {C.~T.}\ \bibnamefont
  {Foxon}},\ }\href {\doibase 10.1209/0295-5075/22/1/011} {\bibfield  {journal}
  {\bibinfo  {journal} {Europhysics Letters (EPL)}\ }\textbf {\bibinfo {volume}
  {22}},\ \bibinfo {pages} {57} (\bibinfo {year} {1993})}\BibitemShut {NoStop}%
\bibitem [{\citenamefont {Dzurak}\ \emph {et~al.}(1993)\citenamefont {Dzurak},
  \citenamefont {Smith}, \citenamefont {Pepper}, \citenamefont {Ritchie},
  \citenamefont {Frost}, \citenamefont {Jones},\ and\ \citenamefont
  {Hasko}}]{dzurak_observation_1993}%
  \BibitemOpen
  \bibfield  {author} {\bibinfo {author} {\bibfnamefont {A.~S.}\ \bibnamefont
  {Dzurak}}, \bibinfo {author} {\bibfnamefont {C.~G.}\ \bibnamefont {Smith}},
  \bibinfo {author} {\bibfnamefont {M.}~\bibnamefont {Pepper}}, \bibinfo
  {author} {\bibfnamefont {D.~A.}\ \bibnamefont {Ritchie}}, \bibinfo {author}
  {\bibfnamefont {J.~E.~F.}\ \bibnamefont {Frost}}, \bibinfo {author}
  {\bibfnamefont {G.~A.~C.}\ \bibnamefont {Jones}}, \ and\ \bibinfo {author}
  {\bibfnamefont {D.~G.}\ \bibnamefont {Hasko}},\ }\href {\doibase
  10.1016/0038-1098(93)90819-9} {\bibfield  {journal} {\bibinfo  {journal}
  {Solid State Commun.}\ }\textbf {\bibinfo {volume} {87}},\ \bibinfo {pages}
  {1145} (\bibinfo {year} {1993})}\BibitemShut {NoStop}%
\bibitem [{\citenamefont {Beenakker}\ and\ \citenamefont
  {Staring}(1992)}]{beenakker_theory_1992}%
  \BibitemOpen
  \bibfield  {author} {\bibinfo {author} {\bibfnamefont {C.~W.~J.}\
  \bibnamefont {Beenakker}}\ and\ \bibinfo {author} {\bibfnamefont {A.~A.~M.}\
  \bibnamefont {Staring}},\ }\href {\doibase 10.1103/PhysRevB.46.9667}
  {\bibfield  {journal} {\bibinfo  {journal} {Phys. Rev. B}\ }\textbf {\bibinfo
  {volume} {46}},\ \bibinfo {pages} {9667} (\bibinfo {year}
  {1992})}\BibitemShut {NoStop}%
\bibitem [{\citenamefont {Erdman}\ \emph {et~al.}(2017)\citenamefont {Erdman},
  \citenamefont {Mazza}, \citenamefont {Bosisio}, \citenamefont {Benenti},
  \citenamefont {Fazio},\ and\ \citenamefont
  {Taddei}}]{erdman_thermoelectric_2017}%
  \BibitemOpen
  \bibfield  {author} {\bibinfo {author} {\bibfnamefont {P.~A.}\ \bibnamefont
  {Erdman}}, \bibinfo {author} {\bibfnamefont {F.}~\bibnamefont {Mazza}},
  \bibinfo {author} {\bibfnamefont {R.}~\bibnamefont {Bosisio}}, \bibinfo
  {author} {\bibfnamefont {G.}~\bibnamefont {Benenti}}, \bibinfo {author}
  {\bibfnamefont {R.}~\bibnamefont {Fazio}}, \ and\ \bibinfo {author}
  {\bibfnamefont {F.}~\bibnamefont {Taddei}},\ }\href {\doibase
  10.1103/PhysRevB.95.245432} {\bibfield  {journal} {\bibinfo  {journal} {Phys.
  Rev. B}\ }\textbf {\bibinfo {volume} {95}},\ \bibinfo {pages} {245432}
  (\bibinfo {year} {2017})}\BibitemShut {NoStop}%
\bibitem [{\citenamefont {Entin-Wohlman}\ \emph {et~al.}(2010)\citenamefont
  {Entin-Wohlman}, \citenamefont {Imry},\ and\ \citenamefont
  {Aharony}}]{entinwWohlman_three-terminal_2010}%
  \BibitemOpen
  \bibfield  {author} {\bibinfo {author} {\bibfnamefont {O.}~\bibnamefont
  {Entin-Wohlman}}, \bibinfo {author} {\bibfnamefont {Y.}~\bibnamefont {Imry}},
  \ and\ \bibinfo {author} {\bibfnamefont {A.}~\bibnamefont {Aharony}},\ }\href
  {\doibase 10.1103/PhysRevB.82.115314} {\bibfield  {journal} {\bibinfo
  {journal} {Phys. Rev. B}\ }\textbf {\bibinfo {volume} {82}},\ \bibinfo
  {pages} {115314} (\bibinfo {year} {2010})}\BibitemShut {NoStop}%
\bibitem [{\citenamefont {S{\'a}nchez}(2018)}]{banff}%
  \BibitemOpen
  \bibfield  {author} {\bibinfo {author} {\bibfnamefont {R.}~\bibnamefont
  {S{\'a}nchez}},\ }\enquote {\bibinfo {title} {Transport out of locally broken
  detailed balance},}\ in\ \href {\doibase 10.1007/978-3-319-76599-0_3} {\emph
  {\bibinfo {booktitle} {Coupled Mathematical Models for Physical and
  Biological Nanoscale Systems and Their Applications}}},\ \bibinfo {editor}
  {edited by\ \bibinfo {editor} {\bibfnamefont {L.~L.}\ \bibnamefont
  {Bonilla}}, \bibinfo {editor} {\bibfnamefont {E.}~\bibnamefont {Kaxiras}}, \
  and\ \bibinfo {editor} {\bibfnamefont {R.}~\bibnamefont {Melnik}}}\ (\bibinfo
   {publisher} {Springer International Publishing},\ \bibinfo {address}
  {Cham},\ \bibinfo {year} {2018})\ pp.\ \bibinfo {pages} {51--64}\BibitemShut
  {NoStop}%
\bibitem [{\citenamefont {Bosisio}\ \emph {et~al.}(2016)\citenamefont
  {Bosisio}, \citenamefont {Fleury}, \citenamefont {Pichard},\ and\
  \citenamefont {Gorini}}]{bosisio_nanowire_2016}%
  \BibitemOpen
  \bibfield  {author} {\bibinfo {author} {\bibfnamefont {R.}~\bibnamefont
  {Bosisio}}, \bibinfo {author} {\bibfnamefont {G.}~\bibnamefont {Fleury}},
  \bibinfo {author} {\bibfnamefont {J.-L.}\ \bibnamefont {Pichard}}, \ and\
  \bibinfo {author} {\bibfnamefont {C.}~\bibnamefont {Gorini}},\ }\href
  {\doibase 10.1103/PhysRevB.93.165404} {\bibfield  {journal} {\bibinfo
  {journal} {Phys. Rev. B}\ }\textbf {\bibinfo {volume} {93}},\ \bibinfo
  {pages} {165404} (\bibinfo {year} {2016})}\BibitemShut {NoStop}%
\bibitem [{\citenamefont {Jiang}\ \emph {et~al.}(2012)\citenamefont {Jiang},
  \citenamefont {Entin-Wohlman},\ and\ \citenamefont
  {Imry}}]{jiang_thermoelectric_2012}%
  \BibitemOpen
  \bibfield  {author} {\bibinfo {author} {\bibfnamefont {J.-H.}\ \bibnamefont
  {Jiang}}, \bibinfo {author} {\bibfnamefont {O.}~\bibnamefont
  {Entin-Wohlman}}, \ and\ \bibinfo {author} {\bibfnamefont {Y.}~\bibnamefont
  {Imry}},\ }\href {\doibase 10.1103/PhysRevB.85.075412} {\bibfield  {journal}
  {\bibinfo  {journal} {Phys. Rev. B}\ }\textbf {\bibinfo {volume} {85}},\
  \bibinfo {pages} {075412} (\bibinfo {year} {2012})}\BibitemShut {NoStop}%
\end{thebibliography}%

\end{document}